\documentclass[twocolumn]{aastex63}

\usepackage{graphicx}
\usepackage{mathtools} 
\usepackage{xcolor}
\usepackage{xspace}
\usepackage{multirow}
\usepackage{float}

\usepackage{hyperref}

\graphicspath{{./}{figs/}}

\def\R{\textit{R}\xspace}
\def\J{\textit{J}\xspace}
\def\H{\textit{H}\xspace}

\def\NtHp{N$_2$H$^+$\xspace}
\def\NtDp{N$_2$D$^+$\xspace}
\def\DCOp{DCO$^+$\xspace}

\def\HtDp{H$_2$D$^+$\xspace}

\def\oHtDpgrd{1$_{10}$--1$_{11}$\xspace}

\def\CetO{C$^{18}$O\xspace}

\def \H{\textit{H}\xspace}
\def \B{\textit{B}\xspace}
\def \I{\textit{I}\xspace}
\def \Q{\textit{Q}\xspace}
\def \U{\textit{U}\xspace}
\def \PA{\textit{PA}\xspace}
\def \mJybm{mJy~beam$^{-1}$\xspace}

\def\arcmin{\mbox{$^{\prime}$}\xspace}
\def\arcsec{\mbox{$^{\prime\prime}$}\xspace}

\renewcommand\micron{\mbox{$\mu$m}\xspace}

\usepackage{soul}

\graphicspath{{./}}

\received{}
\revised{}
\accepted{}

\shorttitle{Magnetic fields of the starless core L\,1512}
\shortauthors{Lin et al.}

\begin{document}

\title{Magnetic fields of the starless core L\,1512
       }
       
\author[0000-0002-6868-4483]{Sheng-Jun Lin}
\affiliation{Academia Sinica Institute of Astronomy and Astrophysics, No. 1, Section 4, Roosevelt Road, Taipei 10617, Taiwan}
\affiliation{Institute of Astronomy, National Tsing Hua University, No. 101, Section 2, Kuang-Fu Road, Hsinchu 30013, Taiwan}
\affiliation{Center for Informatics and Computation in Astronomy, National Tsing Hua University, No. 101, Section 2, Kuang-Fu Road, Hsinchu 30013, Taiwan}
\email{shengjunlin@asiaa.sinica.edu.tw, slai@phys.nthu.edu.tw}

\author[0000-0001-5522-486X]{Shih-Ping Lai}
\affiliation{Institute of Astronomy, National Tsing Hua University, No. 101, Section 2, Kuang-Fu Road, Hsinchu 30013, Taiwan}
\affiliation{Center for Informatics and Computation in Astronomy, National Tsing Hua University, No. 101, Section 2, Kuang-Fu Road, Hsinchu 30013, Taiwan}

\author[0000-0002-8557-3582]{Kate Pattle}
\affiliation{Institute of Astronomy, National Tsing Hua University, No. 101, Section 2, Kuang-Fu Road, Hsinchu 30013, Taiwan}
\affiliation{Department of Physics and Astronomy, University College London, Gower Street, London WC1E 6BT, United Kingdom}

\author[0000-0001-6524-2447]{David Berry}
\affiliation{East Asian Observatory, 660 N. A’oh\={o}k\={u} Place, University Park, Hilo, HI 96720, USA}

\author[0000-0002-9947-4956]{Dan P. Clemens}
\affiliation{Institute for Astrophysical Research, Boston University, 725 Commonwealth Avenue, Boston, MA 02215, USA}

\author[0000-0002-3319-1021]{Laurent Pagani}
\affiliation{LERMA \& UMR8112 du CNRS, Observatoire de Paris, PSL  University, Sorbonne Universit\'es, CNRS, F-75014 Paris, France}

\author[0000-0003-1140-2761]{Derek Ward-Thompson}
\affiliation{Jeremiah Horrocks Institute, University of Central Lancashire, Preston PR1 2HE, UK}

\author[0000-0003-0334-1583]{Travis J. Thieme}
\affiliation{Institute of Astronomy, National Tsing Hua University, No. 101, Section 2, Kuang-Fu Road, Hsinchu 30013, Taiwan}
\affiliation{Center for Informatics and Computation in Astronomy, National Tsing Hua University, No. 101, Section 2, Kuang-Fu Road, Hsinchu 30013, Taiwan}

\author[0000-0001-8516-2532]{Tao-Chung Ching}
\affiliation{Jansky Fellow, National Radio Astronomy Observatory, 1003 Lopezville Road, Socorro, NM 87801, USA}

\begin{abstract}
We present JCMT POL-2 850~\micron dust polarization observations and Mimir \H band stellar polarization observations toward the starless core L\,1512. 
We detect the highly-ordered core-scale magnetic field traced by the POL-2 data, of which the field orientation is consistent with the parsec-scale magnetic fields traced by Planck data, suggesting the large-scale fields thread from the low-density region to the dense core region in this cloud.
The surrounding magnetic field traced by the Mimir data shows a wider variation in the field orientation, suggesting there could be a transition of magnetic field morphology at the envelope scale.
L\,1512 was suggested to be presumably older than 1.4~Myr in a previous study via time-dependent chemical analysis, hinting that the magnetic field could be strong enough to slow the collapse of L\,1512.
In this study, we use the Davis–Chandrasekhar–Fermi method to derive a plane-of-sky magnetic field strength ($B_{\rm pos}$) of 18$\pm$7~$\mu$G and an observed mass-to-flux ratio ($\lambda_{\rm obs}$) of $3.5\pm2.4$, suggesting that L\,1512 is magnetically supercritical.
However, the absence of significant infall motion and the presence of an oscillating envelope are inconsistent with the magnetically supercritical condition. 
Using a Virial analysis, we suggest the presence of a hitherto hidden line-of-sight magnetic field strength of $\sim$27~$\mu$G with a mass-to-flux ratio ($\lambda_{\rm tot}$) of $\sim$1.6, in which case both magnetic and kinetic pressures are important in supporting the L\,1512 core.
On the other hand, L\,1512 may have just reached supercriticality and will collapse at any time.

\end{abstract}
\keywords{Interstellar magnetic fields (845); Interstellar medium (847); Molecular clouds (1072); Polarimetry (1278); Submillimeter astronomy (1647); Star forming regions (1565); Star formation (1569); Starlight polarization (1571)}

\section{Introduction}\label{sec:intro}

Magnetic fields (\B-fields) may play a key role in the formation of starless cores and the evolution of protostellar cores.
In a magnetic-field-dominated scenario, the magnetic field tends to be uniform, and a starless core will collapse into a flattened structure (a so-called pseudo-disk) and pull the magnetic field into an hourglass shape with its symmetry axis aligned with the minor axis of the pseudo-disk and with the rotation/outflow axes \citep[e.g.,][]{Shu87, Galli93a, Galli93b, Li96, McKee07}.
On the other hand, magnetohydrodynamic (MHD) simulations suggest that turbulence can cause the magnetic field to become disordered and chaotic, and either compress the gas to form stars or dissipate gas to suppress star-forming activities \citep[e.g.,][]{Padoan99, MacLow04, Federrath12}.

There are numerous observations of the magnetic field morphology in the protostellar stage.
For scales larger than cores ($\sim$0.1~pc), both hourglass-like patterns and chaotic \B-field patterns are found in the protostellar envelopes of young Class 0 and hot massive cores \citep[e.g.,][]{Girart06, Girart09, Ching17}.
At pseudo-disk scales ($\sim$0.01~pc), \citet{Chapman13} found a positive correlation between the mean \B-field direction and the modeled pseudo-disk symmetry axis in six Class 0 sources that have outflows near the plane of the sky. 
On the other hand, \citet{Hull13} showed that the angle between the \B-field direction and outflow is likely statistically random at smaller scales of $\sim$1000~au in a sample of 16 cores hosting Class 0 or I protostars. 
Hence, in order to understand the origin of the variation in the \B-field morphology, it is crucial to measure dust polarization at the early, starless stage.

Unlike protostellar cores, the magnetic fields of starless cores are rarely measured due to their low polarized thermal emission intensity.
Observations of background starlight polarimetry in the near-infrared (NIR) have been performed toward a few starless cores.
These results showed significant non-uniformity in the plane-of-sky \B-field structure around the cores \citep[e.g.,][]{Clemens16, Kandori17a, Kandori20b, Kandori20c}.
While background starlight polarimetry provides insights into these peripheral regions, it has limitations when it comes to probing the magnetic fields in the large-$A_{\rm V}$ central regions.
To address this, dust continuum emission polarimetry in the submillimeter (submm) wavelengths has emerged as a key tool for investigating the magnetic fields in starless cores. 
However, starless cores, typically extended objects without compact substructures, tend to be entirely resolved out by interferometers \citep{Dunham16, Kirk17, Caselli19, Tokuda20}. As a result, single-dish telescopes, such as the James Clerk Maxwell Telescope (JCMT) and the Atacama Pathfinder Experiment (APEX) telescope, remain the preferred tools for observing starless cores.

Mapping the magnetic field morphology across entire cores is largely restricted to nearby star-forming regions due to low surface brightness and small core sizes.
So far, only ten nearby individual starless cores have had 850~\micron/870~\micron dust polarization detection: 
L\,183, L\,1544, L\,43 \citep{Ward-Thompson00, Crutcher04}, L\,1498, and L\,1517B \citep{Kirk06} were observed with JCMT-SCUPOL;
FeSt\,1-457 \citep{Alves14} with APEX-PolKa; 
and Oph\,C \citep{LiuJH19}, IRAS\,16293E, L\,1689 SMM-16, and L\,1689B \citep{Pattle21} with JCMT-POL2. 
Their core sizes (the full width at half maximum, FWHM, of the submm intensity) range from approximately 0.04~pc to 0.1~pc, for distances spanning from approximately 110~pc to 150~pc \citep{Kirk05, Alves14, LiuJH19, Karoly20, Pattle21}.
With resolutions of 14\arcsec and 20\arcsec for JCMT and APEX, respectively, a spatial resolution of $\sim$0.01~pc can be achieved at these distances to resolve core structure. 
In addition, a number of smaller starless cores embedded in clumps and filaments have also had polarization detection with JCMT-POL2 \citep{Pattle19FrASS}.
However, these studies mostly focused on either the role of \B-fields at the parent clumps themselves \citep[e.g., Oph A \& B clumps;][]{Kwon18, Soam18} or the mean \B-field orientations of the cores compared to large-scale \B-fields \citep[e.g., L\,1495 filament;][]{Eswaraiah21, Ward-Thompson23}, rather than the resolved \B-field morphologies in the individual cores, due to sensitivity limitations.
The protostellar cores have brighter surface brightness and internally complex substructures (e.g., disks, outflows), making them not just easier to be observed with single-dishes \citep[e.g.,][]{Chapman13, Pattle21} but also more commonly observed with interferometers with much higher resolutions \citep[see][and references therein]{Hull19}.

The aforementioned nearby isolated starless cores often exhibit magnetic fields that are relatively smooth and well-ordered \citep{Pattle23_PP7}. Of the first three starless core detected with submm polarizations \citep[L\,183, L\,1544, and L\,43;][]{Ward-Thompson00},
\citet{Crutcher04} smoothed these data from 14\arcsec to 21\arcsec and applied the Davis-Chandrasekhar-Fermi \citep[DCF; ][]{Davis51DCF, Chandrasekhar53DCF} method, but could not conclude whether the cores were magnetically subcritical (i.e., magnetically supported) due to the unknown orientations/inclinations of the cores.  
One unexpected result from the submm single-dish observations is that the mean \B-field direction and the core minor axes are not perfectly aligned.
\citet{Ward-Thompson09} found an angular offset of ${\sim20^\circ\text{--}50^\circ}$ among the five polarization-detected cores \citep{Ward-Thompson00, Kirk06} available at that time.
This is opposite to what magnetically-regulated star formation models predict.
Although this offset may be explained by the projection effect of tri-axial objects \citep{Basu00}, observations with a higher signal-to-noise ratio (SNR) are needed to critically compare to theoretical models. 
Recently, L\,183 and L\,43 were re-observed with JCMT-POL2 \citep{Karoly20, Karoly23} and the results suggest that L\,183 is magnetically subcritical throughout the entire core and L\,43 is also subcritical.
The \citeauthor{Karoly20} studies also demonstrate the sensitivity of JCMT-POL2 is greatly improved compared with the previous polarimeter JCMT-SCUPOL.

Another challenge for starless core polarimetry is that the power-law index $\alpha$ of the polarization-fraction to total-intensity relation ($p\propto I^{-\alpha}$) has been found to be close to 1 \citep[e.g.,][]{Alves15, LiuJH19}, indicating that grain alignment with the magnetic field inside the core could be lost \citep[e.g.,][]{Goodman95, Andersson15}.
Typically, the polarization fraction measurements are debiased with Gaussian noise and filtered with SNR criteria to fit the above single power-law relation in the conventional approach.
However, \citet{Pattle19} found that in cases of low polarized intensity (e.g., starless cores) or partial grain alignment loss ($0<\alpha<1$), a higher SNR threshold in Stokes \I, which discards more data, is necessary to reliably recover the $\alpha$ index (see their Fig.~1).
Below this SNR threshold, the fitted $\alpha$ index will tend to be 1 because the actual polarization signal is weaker than the non-Gaussian-distributed noise, which has an approximate $I^{-1}$ dependence. 
Therefore, without a suitable SNR threshold, the conventional approach (fitting with a single power-law model of $\alpha$) would overestimate the $\alpha$ index.
\citet{Pattle19} found that $\alpha$ can be better estimated with a Ricean noise model without debiasing the polarization; i.e., favoring a non-debiased-polarization to total-intensity relation ($p\arcmin$--\I). 
The authors thus revised the $\alpha$ index of the starless core Oph\,C to 0.6$\sim$0.7 from the value of 1.03 previously found by \citet{LiuJH19}. In FeSt\,1-457, another starless core, $\alpha$ was also revised from 0.92 \citep{Alves15} to a lower value of 0.41 \citep{Kandori20a} using this method.
\citet{WangJW19} also adopted the same noise model but used a Bayesian approach to revise the $\alpha$ index in IC 5146 from 1.03$\sim$1.08 to 0.56.
In addition, an equivalent relation in NIR is $p/A_V\propto A_V^{-\alpha}$ \citep[][]{Andersson15, Pattle19}.
It has been found that $\alpha<1$ in the dense clouds, even with debiased data \citep[see][and references therein]{WangJW17, WangJW19}.
Therefore, dust grains remain aligned at higher densities than previously expected, allowing investigation of the magnetic fields within starless cores.

L\,1512 \citep{Lynds62} is a nearby isolated dark cloud harboring a starless core located near the edge of the Taurus–Auriga molecular cloud complex \citep{Myers83a, Lombardi10_2MASS_III, Launhardt13} at a nominal distance of 140~pc \citep{Kenyon94, Torres09, Roccatagliata20}.
\citet{Lin20} modeled the physical and chemical structure of the L\,1512 core using infrared dust extinction measurements and multi-line observations of \NtHp, \NtDp, \DCOp, \CetO, and \HtDp (\oHtDpgrd) using non-LTE radiative transfer.
With the high-density tracer \NtHp, they found a low central temperature of 8$\pm$1~K, a small three-dimensional isotropic non-thermal velocity dispersion of 0.080~km~s$^{-1}$ (corresponding to a one-dimensional non-thermal velocity dispersion, $\sigma_{v, {\rm NT}}$, of 0.046~km~s$^{-1}$, a non-thermal FWHM linewidth, $\Delta v_{\rm NT}$, of 0.109~km~s$^{-1}$, and a Mach number, $\mathcal{M}=\sigma_{v, {\rm NT}}/c_s$, of 0.27, where $c_s$ is the isothermal sound speed of 0.17~km~s$^{-1}$ at 8~K), and the absence of inward motions.
\citet{Lin20} concluded that L\,1512 is chemically evolved and  older than 1.4~Myr, suggesting that the dominant core formation mechanism could be a slow process, such as ambipolar diffusion \citep[e.g.,][]{Mouschovias91, Tassis04, Mouschovias06}.
In addition, \citet{Falgarone01} found that the transverse velocity gradients (i.e., gradients perpendicular to filaments) change their sign periodically within the CO filaments surrounding L\,1512, and they suggested an MHD instability may be developing in a helical \B-field within the filaments with $\sim$10~$\mu$G at a density of 500~cm$^{-3}$.
They also found inward motions toward the L\,1512 core along a one-parsec-long north-south filament with a low accretion rate of $\sim4\times10^{-6}$~M$_\odot$~yr$^{-1}$, likely mediated by \B-field forces.
These results suggest that magnetic fields in the L\,1512 core could be strong enough to slow its evolution.

In this paper, we aim to examine the role of magnetic fields in L\,1512 by studying the multi-scale \B-field morphology and evaluating the \B-field strength using linear polarization data.
We describe these polarization observations in Sec.~\ref{sec:observation_L1512_Bf}, our reduction recipes in Sec.~\ref{sec:reduction_L1512_Bf}, and present our result and analysis in Sec.~\ref{sec:res_ana}.
In Sec.~\ref{sec:discussion_L1512_Bf}, we discuss the energy budgets of L\,1512. 
Lastly, we summarize our results in Sec.~\ref{sec:conclusion_L1512_Bf}.

\begin{figure*}[ht]
	\centering
    \includegraphics[scale=0.4]{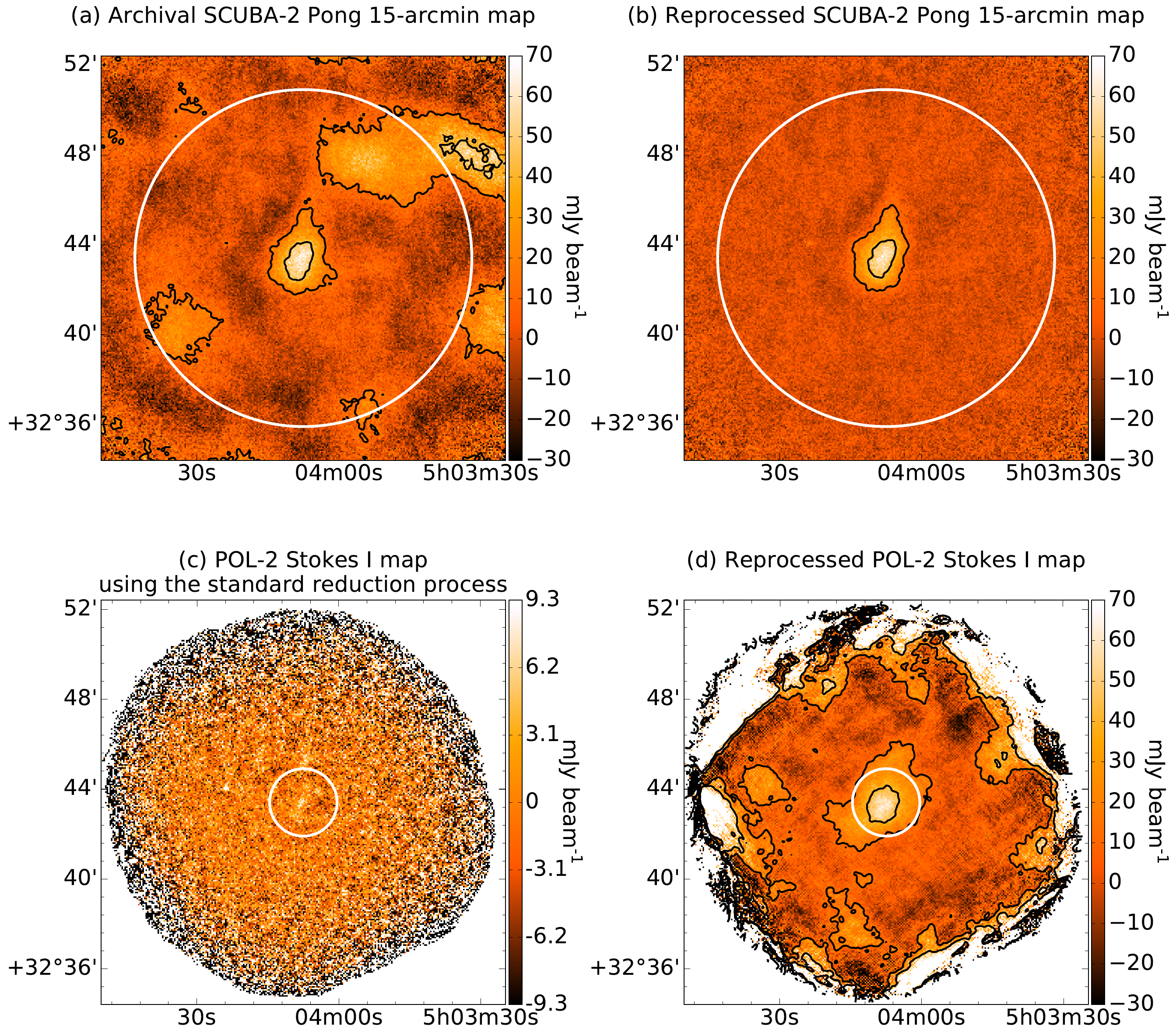}
	\caption{
    L\,1512 JCMT 850~\micron total intensity maps sampled on a 4\arcsec-grid.
    (a) Archival SCUBA-2 Pong-15\arcmin map from the \textit{makemap} routine (see Sec.~\ref{sec:JCMTobs}). 
    (b) Reprocessed SCUBA-2 map from the \textit{skyloop} routine with a PCA model (see Sec.~\ref{sec:red_scuba2}).
    (c) POL-2 Stokes \I map with the standard \textit{pol2map} reduction process, including a PCA model, and 
    (d) the reprocessed POL-2 Stokes \I map from the \textit{makemap} routine with the PCA model disabled (see Sec.~\ref{sec:red_miss}).
    The uniform noise fields with diameters of 15\arcmin for SCUBA-2 and 3\arcmin for POL-2 are indicated by white circles.
    Panels other than (\textit{c}) have the same color scales and same relative contour levels (20\% and 60\%) with respect to their peak intensities (74, 70, and 69~\mJybm for panels (\textit{a}), (\textit{b}), and (\textit{d})) for straightforward comparison.
    The color scale of panel (\textit{c}) is [$-3\sigma$, $3\sigma$], where $\sigma=3.1$~\mJybm is measured in the central 3\arcmin field.
    }
    \label{fig:Imaps}
\end{figure*}

\begin{figure}
	\centering
    \includegraphics[scale=0.5]{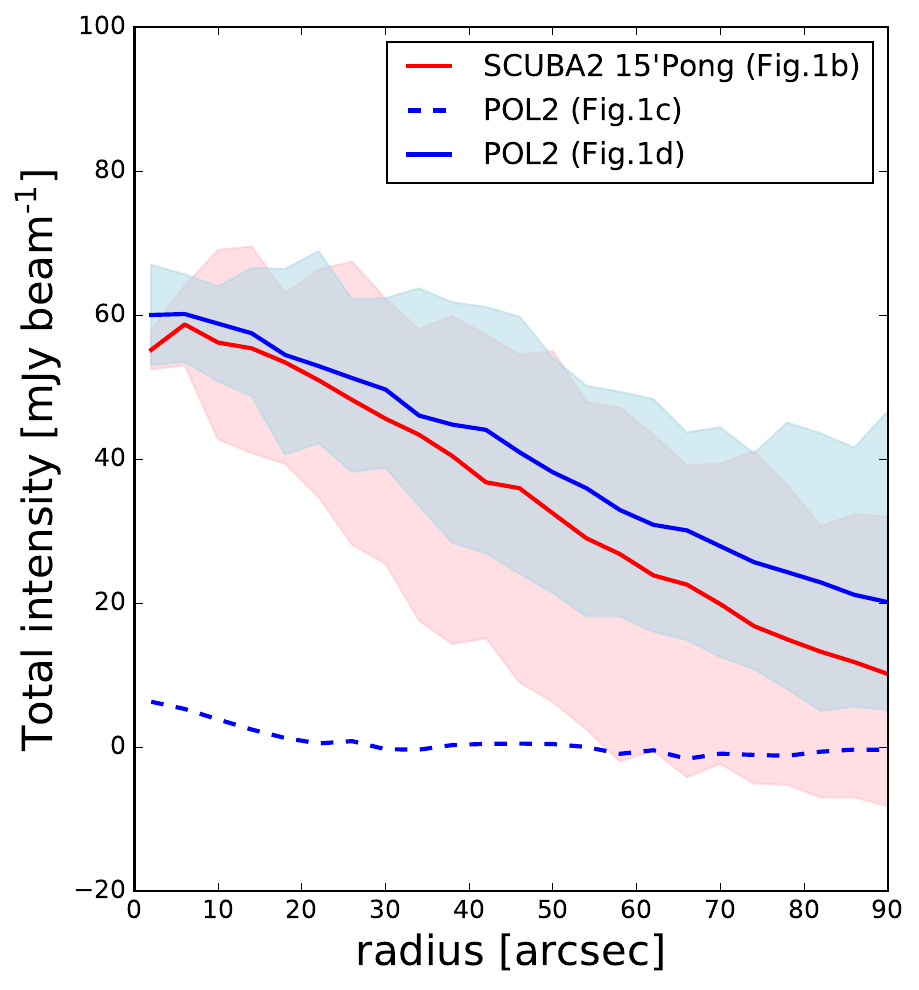}
	\caption{
    L\,1512 850~\micron total intensity profiles.
    The curves are averaged radial profiles while the filled areas show the ranges of the maximum and minimum intensities.
    The red curve and filled area show the profile of the SCUBA-2 map in Fig.~\ref{fig:Imaps}b.
    The blue dashed curve shows the profile of the POL-2 Stokes \I map in Fig.~\ref{fig:Imaps}c.
    The blue solid curve and filled area show the profile of the POL-2 Stokes \I map in Fig.~\ref{fig:Imaps}d.
    }
    \label{fig:Iprofile}
\end{figure}

\begin{figure*}
	\centering
    \includegraphics[scale=0.4]{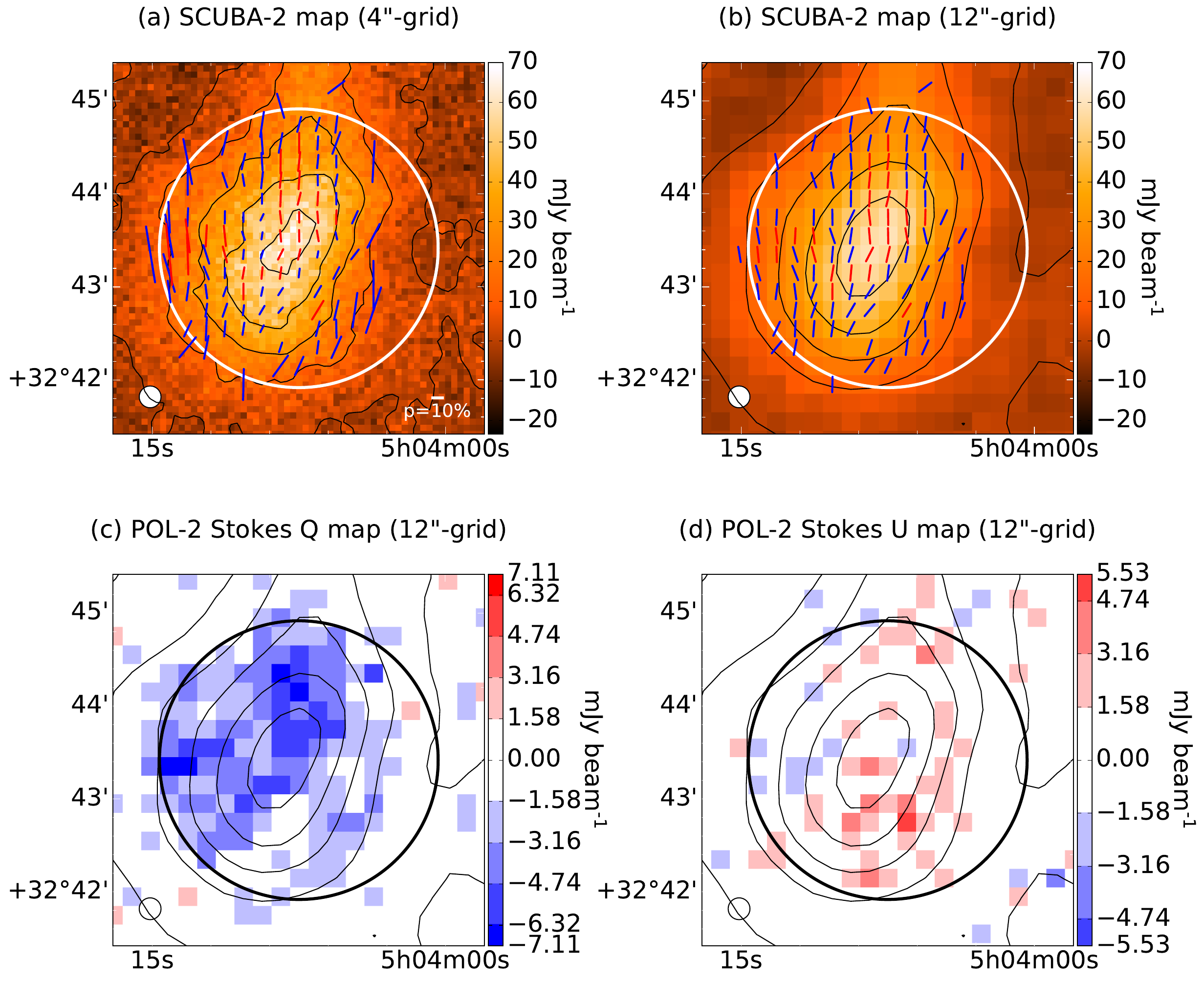}
	\caption{
    L\,1512 JCMT 850~\micron Stokes \I, \Q, and \U maps.
    (a) SCUBA-2 4\arcsec-sampled map (a zoomed-in view of Fig.~\ref{fig:Imaps}b) used as the input total intensity model (see Sec.~\ref{sec:red_miss}) and 
    (b) the same map but smoothed and sampled on a 12\arcsec-grid. Both maps are overlaid with contours at 0\%, 20\%, 40\%, 60\%, and 80\% of their peak intensities. 
    The \B-field vectors (polarization vectors rotated by 90$^\circ$) are overlaid on the 12\arcsec-grid, where vectors with polarization fraction SNR ($p/\sigma_{p}$) larger than 6 are shown in red and vectors with $6> p/\sigma_{p}\geq3$ in blue.
    Vector lengths are scaled according to polarization fraction in panel (\textit{a}), where a scale bar of 10\% (or $p=0.1$ in a decimal equivalent) is denoted in the bottom right corner, and shown with uniform lengths in panel (\textit{b}).
    (c) Stokes \Q and (d) Stokes \U 12\arcsec-sampled maps overlaid with the contours from panel (\textit{b}).
    The beam sizes of 14\arcsec are denoted in the bottom left corners.
    The 3\arcmin POL-2 uniform noise fields are indicated by white and black circles, in which the rms noises are measured as 4.35, 1.03, 0.79, and 0.79~\mJybm in panels (\textit{a})--(\textit{d}), respectively.}
    \label{fig:IQU4}
\end{figure*}

\section{Observations}\label{sec:observation_L1512_Bf}

\subsection{James Clerk Maxwell Telescope (JCMT) Observations}\label{sec:JCMTobs}

The Submillimeter Common-User Bolometer Array 2 \citep[SCUBA-2;][]{Holland13} photometric 850~\micron observations of L\,1512 were retrieved from the JCMT archive\footnote{http://www.cadc-ccda.hia-iha.nrc-cnrc.gc.ca/en/jcmt/}. 
They are part of the M13BC01 project (PI: James di Francesco) observed in November 2013.
The raw data from six observations and one reduced image were downloaded.
Each observation was integrated for 40 minutes with the PONG-900 scan pattern, which fully samples a 15\arcmin diameter circular region.
The telescope beam size at 850~\micron is 14\arcsec.
The downloaded image is presented in Fig.~\ref{fig:Imaps}a, showing some possibly spurious extended structures. 
This image was reduced by the pipeline using the iterative map-making routine \textit{makemap} provided by the \textsc{SMURF} package\footnote{http://starlink.eao.hawaii.edu/docs/sun258.htx/sun258.html} in the \textit{Starlink} software suite \citep{Chapin13}.
We applied another reduction routine on the raw data to eliminate the spurious structures and the details are described in Sec.~\ref{sec:red_scuba2}.

For 850~\micron polarization, we observed L\,1512 with the POL-2 polarimeter \citep{Friberg16} mounted on the SCUBA-2 camera on JCMT.
L\,1512 was observed 27 times, for a total of 14~hours, between August 2020 to November 2020 in Band 2 weather (${0.05<\tau_{\rm 225GHz}<0.08}$) under project code M20BP046 (PI: Sheng-Jun Lin).
Each observation was integrated for 31 minutes with the POL-2-DAISY scan pattern, which fully samples a 3\arcmin diameter circular region.
The reduction details are described in Sec.~\ref{sec:red_miss}.

\subsection{Mimir Polarization Observations}
We conducted \H band polarization observations toward L\,1512 on 2019 December 16, 17, and 20, and on 2020 January 6 and 9, and February 9 with the Mimir instrument \citep{Clemens07} mounted on the 1.8 m Perkins telescope outside Flagstaff, Arizona, U.S., operated by Boston University.
Seeing conditions were between 1.6--3.4~arcsec for all observations.
Data were calibrated with the Basic Data Processor (MSP-BDP) and Photo-Polarimetry Analysis Tool (MSP-PPOL) from the Mimir Software Package\footnote{https://people.bu.edu/clemens/mimir/software.html}.
The data selection criteria were that the \H band magnitude was less than or equal to 13.5~mag, the polarization fraction uncertainty ($\sigma_p$) was less than or equal to 0.9\% (or $\sigma_p\leq0.009$ in a decimal equivalent), and an SNR of the polarization fraction ($p/\sigma_p$) $\geq 2$ or $\sigma_{\PA} \leq 15^\circ$, yielding a sample of 31 stars out of the 194 total observed.

\subsection{Other Observations}

Herschel 500~\micron data with the beam size of 36\arcsec (Observation ID: 1342191182, Quality: level 2 processed) were retrieved from the NASA/IPAC Infrared Science Archive\footnote{https://irsa.ipac.caltech.edu/data/Herschel/HHLI/index.html}.
Planck 353 GHz (850~\micron) polarization data were retrieved from the Planck Legacy Archive\footnote{http://pla.esac.esa.int/pla/\#maps}.
We employed the Planck 2018 \texttt{GNLIC} Stokes \I, \Q, and \U foreground thermal dust emission maps at variable resolution (5\arcmin--80\arcmin) over the sky \citep[PR3;][]{Planck_IV_2020, Planck_XII_2020}, where the Galatic foreground emission is estimated by the \texttt{GNLIC} foreground dust model \citep{Remazeilles11}, with reduced contamination from the cosmic infrared background, cosmic microwave background, and instrumental noise.
The dataset we used for L\,1512 has an effective beam size of 15\arcmin.

\section{Data Reduction}\label{sec:reduction_L1512_Bf}

We describe our JCMT data reduction recipes and compare them to the standard processes in the following.

\subsection{JCMT SCUBA-2 Photometric Map}\label{sec:red_scuba2}

Figure~\ref{fig:Imaps}a shows the archival SCUBA-2 image with the possibly spurious extended structures (see Sec.~\ref{sec:JCMTobs}).
In order to improve the map quality, we reduced the raw SCUBA-2 data using the \textit{skyloop} routine with a configuration file optimized for extended emission\footnote{https://www.eaobservatory.org/jcmt/2019/04/a-new-dimmconfig-that-uses-pca/}, provided by the
\textsc{SMURF} package.
With this configuration file, the \textit{skyloop} routine performed principal component analysis (PCA) to remove correlated noise \citep[see][]{Chapin13}.
Figure~\ref{fig:Imaps}b shows this reprocessed map, where the spurious extended structures are now absent.
The map was gridded to 4\arcsec pixels and calibrated using a flux conversion factor (FCF) of 537~Jy~pW$^{-1}$ \citep{Dempsey13}.

\subsection{Missing Flux of JCMT POL-2 for Extended Structures}\label{sec:red_miss}

The POL-2 polarization data were first reduced with the standard two-stage reduction process\footnote{http://starlink.eao.hawaii.edu/docs/sc22.htx/sc22.html} using the \textit{pol2map} routine in the \textsc{SMURF} package, which includes a PCA model to remove correlated noise.
\citet{Pattle21} contains a detailed description of the standard POL-2 reduction process.
Figure~\ref{fig:Imaps}c presents the POL-2 total intensity map reduced with the above standard procedure, while Figure~\ref{fig:Imaps}d presents the total POL-2 intensity map reduced from the same POL-2 data but with a different reduction procedure that excluded the default PCA model (described later in this section).
In contrast to the obvious detection of L\,1512 in the SCUBA-2 map (Fig.~\ref{fig:Imaps}b, peak at 70~\mJybm), we found that the dust emission of L\,1512 is undetected in Stokes \I (Fig.~\ref{fig:Imaps}c) and neither \Q nor \U are detected with the standard POL-2 data reduction.
Given the rms noise of 3.1~\mJybm measured in the central 3 arcmin field in the POL-2 map using the standard reduction, the dust emission peak of L\,1512 should be detected with an SNR of $\sim$20 based on the SCUBA-2 map.
Therefore, we suspected that the POL-2 scanning pattern or other instrument/reduction issues might have filtered out extended emission when observing low-surface-brightness objects.

One difference between observations with and without POL-2 is the scan speed.
POL-2 contains a half-wave plate (HWP) and a grid analyzer to measure polarized light.
The standard POL-2 observing mode (POL-2-DAISY) is a ``scan and spin" mode, in which the telescope is continuously moving while the HWP spins \citep{Friberg16}.
The POL-2-DAISY scan speed is slow (8\arcsec~s$^{-1}$), which is limited by the integration time for a full polarization cycle (i.e., the HWP rotation speed of 2~Hz). 
Thus the data can be sufficiently sampled and the Stokes parameters \Q and \U can be registered to the maps with at least 4\arcsec pixels, which is the default pixel size (about one-third of the 850~\micron beam size of 14\arcsec; i.e., over-sampled by a factor of $\sim$1.5 relative to a standard Nyquist sampling rate) adopted by the \textit{pol2map} routine.
Such a slow speed could result in less extended structure recovery because the sensitivity of bolometers is limited by the low-frequency noise dominated by slow atmospheric variation between bolometer readouts \citep{Chapin13, Friberg16}.

One approach to evaluate this possibility would be to work with the observatory to perform POL-2 observations with the other scan patterns.
An alternate approach is to disable the PCA model included in the standard two-stage reduction process.
PCA can remove the correlated components of the bolometer readouts. 
However, it may be not a good solution for producing the extended structures because the real signal could be also correlated among the bolometers and hence is removed, as discussed by \citet{Chapin13}.
The POL-2 map size is about half the size of the SCUBA-2 PONG-900 map, so the relative coverage of extended structures on the POL-2 map is larger than that of the SCUBA-2 map. 
As a result, these extended structures are sampled more with POL-2, producing more correlated signals among the bolometers for POL-2. This may explain why the PCA model can help to improve the SCUBA-2 map (Fig.~\ref{fig:Imaps}b) but failed for the POL-2 map (Fig.~\ref{fig:Imaps}c).
We re-reduced the POL-2 raw data using the \textit{makemap} routine without PCA and set \texttt{flagslow=0.01} to take into account the slow scan speed used by POL-2 so that the data taken with the POL-2 scan speed will not be flagged and ignored when producing the map.
This allows comparing the results of the standard two-stage process with and without PCA, to see what the impact of the PCA model is. The new reduction result is shown in Fig.~\ref{fig:Imaps}d.
Without the default PCA model, L\,1512 appears in the new, but noisier, map with a similar peak intensity as seen in our reprocessed SCUBA-2 map (Fig.~\ref{fig:Imaps}b).

Figure~\ref{fig:Iprofile} shows the azimuthally-averaged intensity profiles of our improved SCUBA-2 map (Fig.~\ref{fig:Imaps}b) compared to the (no PCA) POL-2 total intensity map (Fig.~\ref{fig:Imaps}d).
The intensity profiles are similar, showing evidence that we did detect the signal from L\,1512 with POL-2, although we cannot rule out the possibility that parts of the emission were still filtered out.
The standard POL-2 data reduction procedure incorporates the PCA model for the purpose of removing background signals. This step is particularly essential due to the non-uniform and radially dependent background variation within the POL-2 observing field.
Given that all bolometers in POL-2 observe the same background but with time delays in the time-stream data, these signals are correlated and  identified as such by PCA.
The PCA approach does effectively address the background signal issue.
However, the presence of extended structures also leads to correlated total intensity signals between bolometer readouts, and these are subsequently removed by the PCA model.
The above experiment of not using the PCA model indicates that the Stokes \I intensity from L\,1512 could be recovered in the polarization data despite the limitations in map quality.

Therefore, the POL-2 missing flux issue in the total intensity of L\,1512 is likely due to several factors: (1) The scan speed of 8\arcsec~s$^{-1}$ is slow, (2) The background variation is radially dependent in the POL-2 observations, and (3) L\,1512 is faint and extended, resulting in a low surface brightness contrast.

On the other hand, the HWP rotation speed of 2 Hz provides a fast 8 Hz modulation of linear polarized intensity, making the signals of Stokes \Q and \U extractable \citep{Friberg16}. 
However, POL-2 data reduction requires a total intensity (Stokes \I) map to define the source region with a fixed SNR to enable reducing the Stokes \Q and \U signals.
We thus used the reprocessed SCUBA-2 map (Fig.~\ref{fig:Imaps}b and Fig.~\ref{fig:IQU4}a for a zoomed-in view) as the input Stokes \I model for the POL-2 data reduction instead of the POL-2 Stokes \I map.
The reduced \Q and \U maps were gridded to 4\arcsec pixels and calibrated using a POL-2 FCF of 725~Jy~pW$^{-1}$ \citep{Dempsey13, Friberg16}.
This POL-2 data reduction, involving the use of an external SCUBA-2 map, was recommended upon the initial release of POL-2 for open use \citep{Friberg16}.
This procedure was employed in the first POL-2 850~\micron polarization studies by \citet{Ward-Thompson17, Pattle17}, which focused on the Orion A filament, marking the initial publication from the B-fields In STar-forming Region Observations (BISTRO) survey.
The newer standard procedure, utilizing the POL-2 Stokes \I map, was adopted in subsequent BISTRO publications shortly after \citet{Kwon18}, as the targeted objects typically exhibit notable brightness, thereby facilitating effective reduction based on the POL-2 Stokes \I data.

Figure~\ref{fig:IQU4} shows the Stokes \I, \U, and \Q maps, where the Stokes \I maps are overlaid with \B-field vectors.
Panel (a) shows a zoomed-in view of 4\arcsec-sampled Fig.~\ref{fig:Imaps}b, while panels (b), (c), and (d) are binned and sampled on 12\arcsec-grid.
The polarization vectors were computed from the Stokes \I, \Q, and \U maps on the 12\arcsec-grid to make each vector a nearly independent measurement (close to the resolution of 14\arcsec).
Assuming dust grains are aligned with the magnetic field, the dust emission polarization at submm wavelengths is perpendicular to the plane-of-sky magnetic field, while the dust-extincted starlight polarization in the NIR is parallel to the plane-of-sky magnetic field \citep[e.g.,][]{Andersson15, Pattle23_PP7}.
Thus the vectors overlaid on Fig.~\ref{fig:IQU4}a and \ref{fig:IQU4}b were rotated 90$^\circ$ so they may represent the plane-of-sky \B-field direction. 
The vectors have been then filtered, based on the SNR of the polarization fraction being larger than three ($p/\sigma_{p}\geq3$) and the SNR of the total intensity larger than ten ($I/\sigma_{I}\geq10$).
We note that a few vectors appear in the periphery with very high polarization fractions of about 50\%.
This is probably because the total intensity is underestimated, as the whole point of using a SCUBA-2 PONG observation is to maximize large-scale structure recovery, but the flux could still be missed, as L\,1512 is fainter and more extended (i.e., low-contrast surface brightness) compared to the other  starless cores which have had 850~\micron polarization detection (see Sec.~\ref{sec:intro}).
The average rms noise of the Stokes \Q and \U maps over the central 3\arcmin field is $\sim$0.79 mJy for 12\arcsec pixels. For the total intensity map, the average rms noise is $\sim$1.03 mJy for 12\arcsec pixels.
We note that the total intensity rms noise is not related to the Stokes \Q and \U rms noise because we used the total intensity map observed with SCUBA-2 instead of POL-2.

\begin{figure}
	\centering
    \includegraphics[scale=0.5]{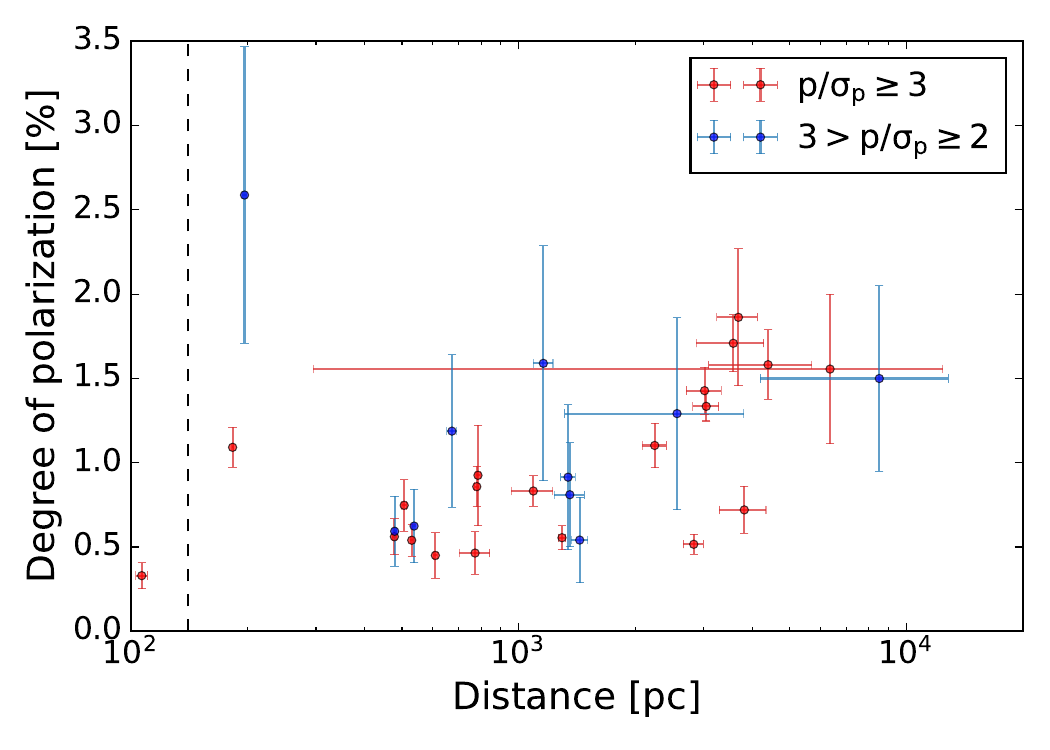}
	\caption{
    Mimir \H band polarization fraction ($p$), expressed in percentage, is plotted against the stellar distance for 30 stars with positive parallax measurements ($\pi$) from Gaia DR3 \citep{Gaia_DR3_2023}, out of the total 31 stars. Dots with their uncertainties are shown in red for $p/\sigma_{p}\geq3$, and in blue for $3>p/\sigma_{p}\geq2$. The black dashed line indicates the nominal distance to L\,1512 of 140~pc.}
    \label{fig:Gaia}
\end{figure}

\begin{figure*}
	\centering
    \includegraphics[scale=0.45]{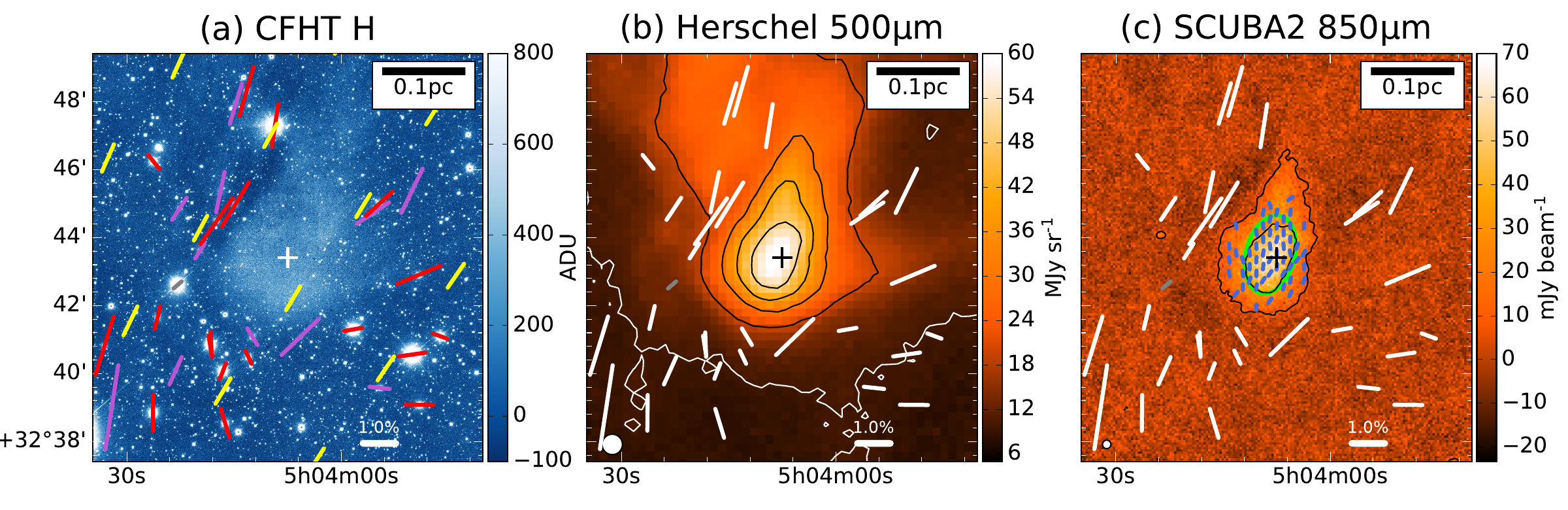}
	\caption{
	L\,1512 Mimir \H band background starlight polarimetry compared to dust maps.
    Mimir polarizations are shown as red, magenta, or white vectors on each panel with scale bars of 1\% polarization fraction (or $p=0.01$ in a decimal equivalent).
    One star shows a grey vector as it is a foreground star, identified using Gaia data in Fig.~\ref{fig:Gaia}. 
    (a) CFHT \H band image from \citet{Lin20} and yellow Planck 353 GHz \B-field vectors.
    The Mimir vectors are displayed in magenta for $3>p/\sigma_{p}\geq2$ and in red for $p/\sigma_{p}\geq3$.
    (b) Herschel 500~\micron map with contours at 10, 20, 30, 40, and 50 MJy sr$^{-1}$.
    (c) JCMT SCUBA-2 850~\micron map from Fig.~\ref{fig:Imaps}b with contours at 10\%, and 50\% of the peak intensity at 69.5 mJy beam$^{-1}$.
    The 2D Gaussian intensity fitting result is denoted by a green ellipse.
    The JCMT 850~\micron \B-field vectors with $p/\sigma_{p}>3$ are displayed in blue with uniform lengths.
    The central cross in each panel indicates the center of L\,1512 \citep{Lin20}.
    The scale bars of 0.1~pc and submillimeter-wavelength beam sizes are denoted in the top right and bottom left corners, respectively.}
    \label{fig:NIR}
\end{figure*}

\subsection{Polarization Properties}
The formulae for computing the polarization properties from the reduced data are as follows:
the non-debiased polarization fraction ($p\arcmin$) is given by 
\begin{equation}
p\arcmin=\frac{1}{I}\sqrt{Q^2 + U^2},
\label{equ:p_prime}
\end{equation}
where \I, \Q, \U are the Stokes parameters measured. 
We let $\sigma_I$, $\sigma_Q$, and $\sigma_U$ be their uncertainties.
Uncertainty of the polarization fraction ($\sigma_p$) is given by
\begin{equation}
\sigma_p=\sqrt{\frac{Q^2\sigma_Q^2+U^2\sigma_U^2}{I^2(Q^2+U^2)}+\frac{(Q^2+U^2)\sigma_I^2}{I^4}}.
\label{equ:sig_p}
\end{equation}
The polarization fraction is debiased using the asymptotic estimator \citep{Vaillancourt06, Montier15}, 
\begin{equation}
p=\sqrt{p\arcmin^2 - \sigma_p^2}\label{equ:p},
\end{equation}
and is rearranged by adopting the aforementioned ${\sigma_Q = \sigma_U = 0.79}$~\mJybm and ignoring the $I^{-4}$ term of the radicand in Equation~\ref{equ:sig_p} \citep{Kwon18} as
\begin{equation}
p\approx \frac{1}{I}\sqrt{Q^2 + U^2 - \frac{1}{2}(\sigma_Q^2+\sigma_U^2)},
\end{equation}
where $p$ is the debiased polarization fraction. 
The polarization angle (\PA) is given by
\begin{equation}
\PA=\frac{1}{2}\arctan{\left(\frac{U}{Q}\right)},
\end{equation}
and the uncertainty of the polarization angle ($\sigma_{\PA}$) by 
\begin{equation}
\sigma_{PA}=\frac{1}{2}\frac{\sqrt{Q^2\sigma_U^2 + U^2\sigma_Q^2}}{Q^2 + U^2}.
\end{equation}

\begin{figure}
	\centering
    \includegraphics[scale=0.8]{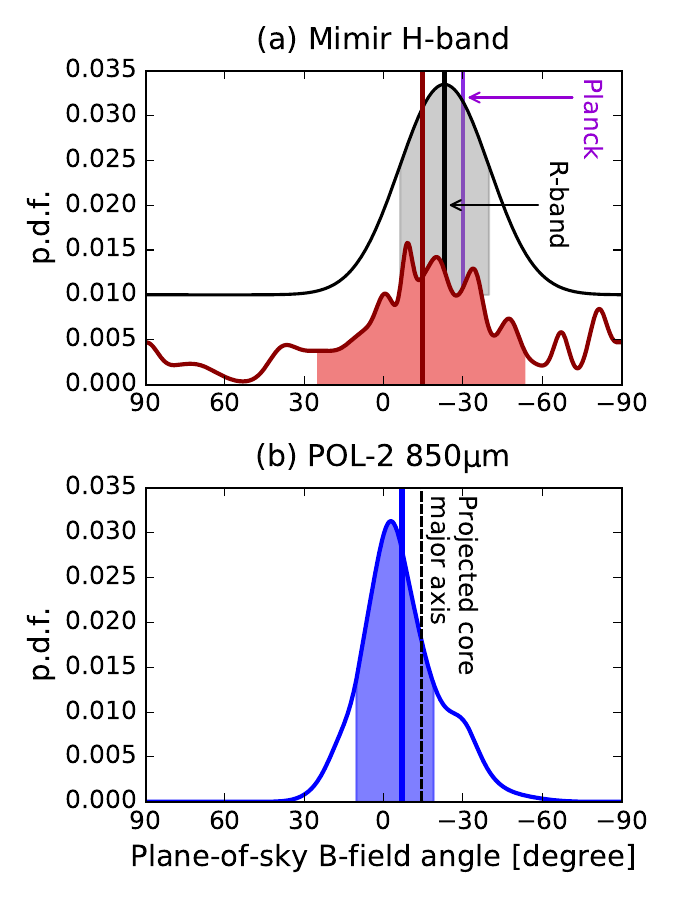}
	\caption{
    Probability density function (p.d.f.) of \B-field position angles ($\theta$).
	The upper panel shows the Mimir data (red), while the lower panel shows the JCMT POL-2 data (blue).
	The mean and the 68\% highest density interval (HDI) of the \B-field angle distributions are displayed with vertical lines and filled areas.
    The Planck mean field angle of $-$30$^\circ$ is plotted as a purple line in panel (\textit{a}).
    The mean field angle, with standard deviation, of AIMPOL R-band polarizations \cite{Sharma22} is represented as a Gaussian distribution on panel (\textit{a}), vertically offset by a probability of 1\% for clarity.
    The position angle of the core major axis ($\PA=-14.2^\circ$; see Fig.~\ref{fig:NIR}c) resulting from the 2D Gaussian fitting to the 850~\micron intensity is plotted as a black dashed line in panel (\textit{b}).}
    \label{fig:PA_hist}
\end{figure}

\section{Results and analysis}\label{sec:res_ana}

\subsection{Foreground Star Census}\label{sec:foreground}

The dust-extincted starlight polarization in the Mimir \H band from the background stars can trace the plane-of-sky magnetic field toward L\,1512.
We aim to identify and exclude foreground stars from our \H band polarimetric detection.
We match our 31 Mimir NIR stellar polarization measurements with the Gaia parallax ($\pi$) measurements \citep[DR3;][]{Gaia_DR3_2023}.
Figure~\ref{fig:Gaia} shows the \H band polarization fraction ($p$) plotted against their stellar distance, except for one star with a negative parallax ($-0.0480\pm0.2224$~mas). 
The stellar distance ($d$) is calculated by, $d=1/\pi$, and the corresponding uncertainty is approximated by $\sigma_{d}=\sigma_\pi/\pi^2$ \citep{Luri18}.
We adopt the nominal distance of 140~pc to L\,1512 \citep{Kenyon94, Torres09, Roccatagliata20} and identify one foreground star with a distance of $107\pm4$~pc. This foreground star indeed has the lowest \H band polarization fraction among our data ($0.330\pm0.077$\%, expressed in percentage).
The other 30 stars are categorized as background stars due to their distances being greater than $183\pm1$~pc.
Accordingly, they trace the magnetic field toward L\,1512 because CO isotopologue line observations exhibit only one gas component along the sightlines \citep{Falgarone98}.
Figure~\ref{fig:NIR}a shows the Mimir \H band polarization vectors superimposed on the CFHT \H band image. A grey vector corresponds to the foreground star, while the remaining 30 vectors are associated with background stars.
The sightlines toward these background stars sample from about 130\arcsec away from the core center \citep[$\text{R.A.} = 5^{\mathrm h}04^{\mathrm m}07\fs5$ and $\text{decl.} = 32\arcdeg43\arcmin25\farcs0$, J2000;][]{Lin20} and continue out to 410\arcsec.
Therefore, their polarization angles mostly trace the morphology of the plane-of-sky magnetic field in the diffuse envelope, spanning from a spatial scale of $\sim$0.56~pc down to $\sim$0.18~pc, which is the outer layers of the L\,1512 cloud surrounding the dense core.

\subsection{Magnetic Field Morphology}\label{sec:Bfield}

Figure~\ref{fig:NIR} shows Mimir \H band polarization (\B-field) vectors overlaid on the continuum maps.
These images trace different spatial structures of L\,1512.
Both the JCMT 850~\micron and Herschel 500~\micron dust emission trace the central cold dust.
Herschel had much greater sensitivity to large-scale emission, while SCUBA-2 mostly traces the cold dust in the L\,1512 core.
The \H band dust scattered light \citep[cloudshine; ][]{Foster06} traces the L\,1512 envelope \citep[the outer diffuse region surrounding the L\,1512 core;][]{Saajasto21}.

The Mimir vectors mostly trace the magnetic field morphology in the diffuse envelope, with a spatial scale of $\sim$0.56~pc down to $\sim$0.18~pc (see Sec.~\ref{sec:foreground}).
The envelope-scale \B-field pattern seems to generally follow the large-scale magnetic field traced by the Planck 353 GHz (850~\micron) data.
These Planck vectors were rotated 90$^\circ$ to represent the plane-of-sky \B-field directions ($\theta=\PA + 90^\circ$) and over-sampled for visualization, as the entire field of view (11.5\arcmin$\times$12\arcmin) of Fig.~\ref{fig:NIR}a fits within one effective Planck beam size of 15\arcmin. 
The Planck data indicate the large-scale mean field angle of $\theta_{\rm Planck}=-30^\circ$ over a spatial scale of $\sim$0.6~pc at the distance of 140~pc.

In the southern region of L\,1512, there are significant deviations of the NIR vector orientations compared to the Planck vector orientations.
For these NIR vectors, the absolute deviation of their polarization angles around the Planck mean field angle (i.e., $|\theta_{\rm H}-\theta_{\rm Planck}|$) ranges from a minimum of $6^\circ\pm13^\circ$ to a maximum of $80^\circ\pm8^\circ$, with
the 68th percentile absolute deviation at 51$^\circ$.
In contrast, the northern NIR vectors show comparatively less deviation from the Planck mean field angle, with the 68th percentile absolute deviation at 18$^\circ$.
This indicates a \B-field orientation transition from large- to envelope-scales. 
These deviations of the NIR vectors seem to show the effects of a relative motion between the surrounding medium and L\,1512, and most of the interaction with the surrounding medium occurs in the south-southeast.

In highly extincted regions, NIR background starlight is faint and the corresponding NIR polarization is difficult to detect.
To investigate the magnetic fields in the core, polarimetry at longer, submillimeter wavelengths is necessary. 
In Sec.~\ref{sec:red_miss}, Figure~\ref{fig:IQU4}b has shown the JCMT POL-2 \B-field vectors overlaid on the 850~\micron continuum map, zoomed to the core region.
Unlike the envelope-scale \B-field, the core-scale \B-field morphology in L\,1512 shows a much more ordered pattern in a nearly vertical orientation ($\theta_{\rm POL2} \approx 0^\circ$).
We note that the pattern exhibits a mostly smooth change in position angle from $\theta_{\rm POL2} \approx 0^\circ$  in the northwest to $\approx$~$-30^\circ$ in the southeast, similar to the large-scale mean field angle of $-30^\circ$ seen in the Planck data.
The POL-2 data may reveal a twist or kink in the plane-of-sky orientations in the core region, blended with the large-scale magnetic field.
Figure~\ref{fig:NIR}c shows a spatial comparison between POL-2 vectors and Mimir vectors,
where the POL-2 vectors are identical to those in Fig.~\ref{fig:IQU4}b but undersampled by a factor of two for clarity. 
Their distinct different spatial coverages indeed show that the Mimir data trace the envelope-scale \B-field while the POL-2 data trace the core-scale \B-field.

To further compare the magnetic field orientations at different scales, Figure~\ref{fig:PA_hist} shows the distributions of the plane-of-sky \B-field position angle ($\theta$) estimated from our Mimir and POL-2 data plus the mean field angle inferred from the Planck data. 
These \B-field angle probability density functions (p.d.f.) are derived to account for the uncertainties ($\sigma_{\PA}$) in the individual \PA measurements via accumulation of the representative Gaussian distributions \citep{Clemens20}. 
For POL-2 and Mimir data, the mean \B-field angle and the 68\% highest density interval (68\% HDI; equivalent to one standard deviation of a Gaussian distribution) calculated from the p.d.f. are
${-7^\circ}_{-12^\circ}^{+18^\circ}$ and 
${-15^\circ}_{-39^\circ}^{+40^\circ}$, respectively.
In addition, the mean \B-field angle with standard deviation of $-23^\circ\pm17^\circ$ inferred from the AIMPOL \R band polarization of 94 stars \citep{Sharma22} is also overplotted.
The optical \textit{R} band polarization traces the magnetic field beyond the region probed by our NIR \H band polarization because the dust extincts more in the \textit{R} band.

Both the AIMPOL data and Planck data trace the large-scale plane-of-sky magnetic field of L\,1512, but with different resolutions. 
The Planck measurements, with their low spatial resolution of 15\arcmin and small optical depth at 850~\micron \citep{Planck_XII_2020}, not only average the polarization on a large scale toward the L\,1512 cloud but also integrate it along the line of sight, weighting it by dust emission intensity.
In contrast, the \R band measurements could not be made toward the central $\sim$10\arcmin extincted area of the L\,1512 cloud \citep[see Fig.~4 of][]{Sharma22}. However, the AIMPOL data coverage has a diameter of $\sim$20\arcmin ($\sim$0.8~pc at the distance of 140~pc).
As a result, the magnetic field traced in \R band spans a wide field, sampling primarily the outer low-density region, but with relatively higher resolution, depending on the \R band background star distribution.
Given that foreground and background stars can be identified, one can confidently associate the measured polarization with the cloud.
Consequently, both the Planck and AIMPOL measurements effectively trace the large-scale magnetic field pattern and are consistent with each other toward the low-density periphery of the L\,1512 cloud.

The magnetic field toward L\,1512 shows a consistent average orientation from the large scale down to the core scale. 
The largest angular dispersion is seen in the \H band data, related to the previously mentioned deviation of the NIR vectors in the southern region.
In addition, \citet{Sharma22} found that the \textit{R} band polarization fraction is lower toward the southern region compared to the other surrounding regions. 
They suggested that it could be due to the lack of aligned grains owing to the different grain size distribution, collisional disalignment with gas, or the depolarization caused by a tangled \B-field.
Therefore, with these multiwavelength polarimetric data, magnetic fields are suggested to thread from the large scale to the dense core scale in L\,1512, while the magnetic field could interact with the diffuse medium in the envelope scale.

The dust emission of L\,1512 (Figs.~\ref{fig:NIR}b and \ref{fig:NIR}c) shows an elongated, cometary morphology. 
We performed a 2D Gaussian intensity fitting on the 850~\micron map. 
The fitting result is denoted by the green ellipse in Fig.~\ref{fig:NIR}c.
The PA of the major axis is ${-14.2^\circ\pm0.4^\circ}$, and the major and minor FWHM axes are ${141.2\arcsec\pm2.1\arcsec}$ and ${86.2\arcsec\pm1.3\arcsec}$ ($\sim$0.10~pc and $\sim$0.06~pc at the distance of 140~pc), respectively. 
The major axis of the projected core shape is parallel to the \B-field orientation (see Fig.~\ref{fig:PA_hist}b) instead of being perpendicular to the magnetic field as suggested by the \B-field-dominated core formation scenario at the core scale of $\sim$0.1~pc \citep[e.g.,][]{Galli93a, Li96, Ciolek00, Myers18}.
This discrepancy can be attributed to the projection effect of a tri-axial core, as suggested by \citet{Basu00}. \citet{ChenCY18} conducted a comprehensive analysis of turbulent MHD simulations and they discovered that dense cores do tend to be tri-axial, unlike the idealized oblate cores assumed in classical theories. 
Moreover, they observed that environmental factors, such as ram pressure or magnetic pressure, also play a crucial role in shaping dense cores.

\subsection{Davis–Chandrasekhar–Fermi Analysis}

The Davis–Chandrasekhar–Fermi \citep[DCF;][]{Davis51DCF, Chandrasekhar53DCF} method is widely used to derive the plane-of-sky \B-field strength ($B_{\rm pos}$) using linear polarization data.
This method assumes that the turbulence is sub-Alfv\'{e}nic and the magnetic field is frozen into the gas, so the non-thermal gas motions result in a distortion of the magnetic field.
By measuring the dispersions in the non-thermal gas velocities and in the polarization position angles, the field strength is \citep{Crutcher04}:
\begin{align}
B_{\rm pos}&=\mathrm{Q}\sqrt{4 \pi \langle \rho \rangle}\frac{\sigma_{v, {\rm NT}}}{\delta \theta}\,\,{\rm (cgs\,\,units)}\\
&\approx
9.3 \sqrt{\langle n_{\rm H_2} \rangle/{\rm cm}^{-3}} \frac{\Delta v_{\rm NT}/{\rm km\cdot s^{-1}}}{\delta \theta/{\rm degree}} \,\,\mu G,
\label{equ:DCF}
\end{align}
where $\langle \rho\rangle$ is the mean volume mass density, 
$\langle n_{\rm H_2}\rangle$ is the mean H$_2$ number density 
($\langle \rho\rangle$ = $\mu_{\rm H_2} \langle n_{\rm H_2}\rangle  m_{\rm H}$, and
$\mu_{\rm H_2}$ = 2.8),
$\sigma_{v, {\rm NT}}$ is the 1D non-thermal velocity dispersion,
$\Delta v_{\rm NT}$ is the non-thermal FWHM linewidth
($\Delta v_{\rm NT} = \sigma_{v, {\rm NT}} \sqrt{8\log{2}}$), 
$\delta \theta$ is the intrinsic dispersion in \B-field angles,
and $\mathrm{Q}$ is a correction factor.
Based on calibrations from numerical simulations \citep{Heitsch01, Ostriker01}, $\mathrm{Q}=0.5$ provides a good estimate of $B_{\rm pos}$, if $\delta \theta < 25^\circ$.

\begin{figure*}
	\centering
    \includegraphics[scale=0.35]{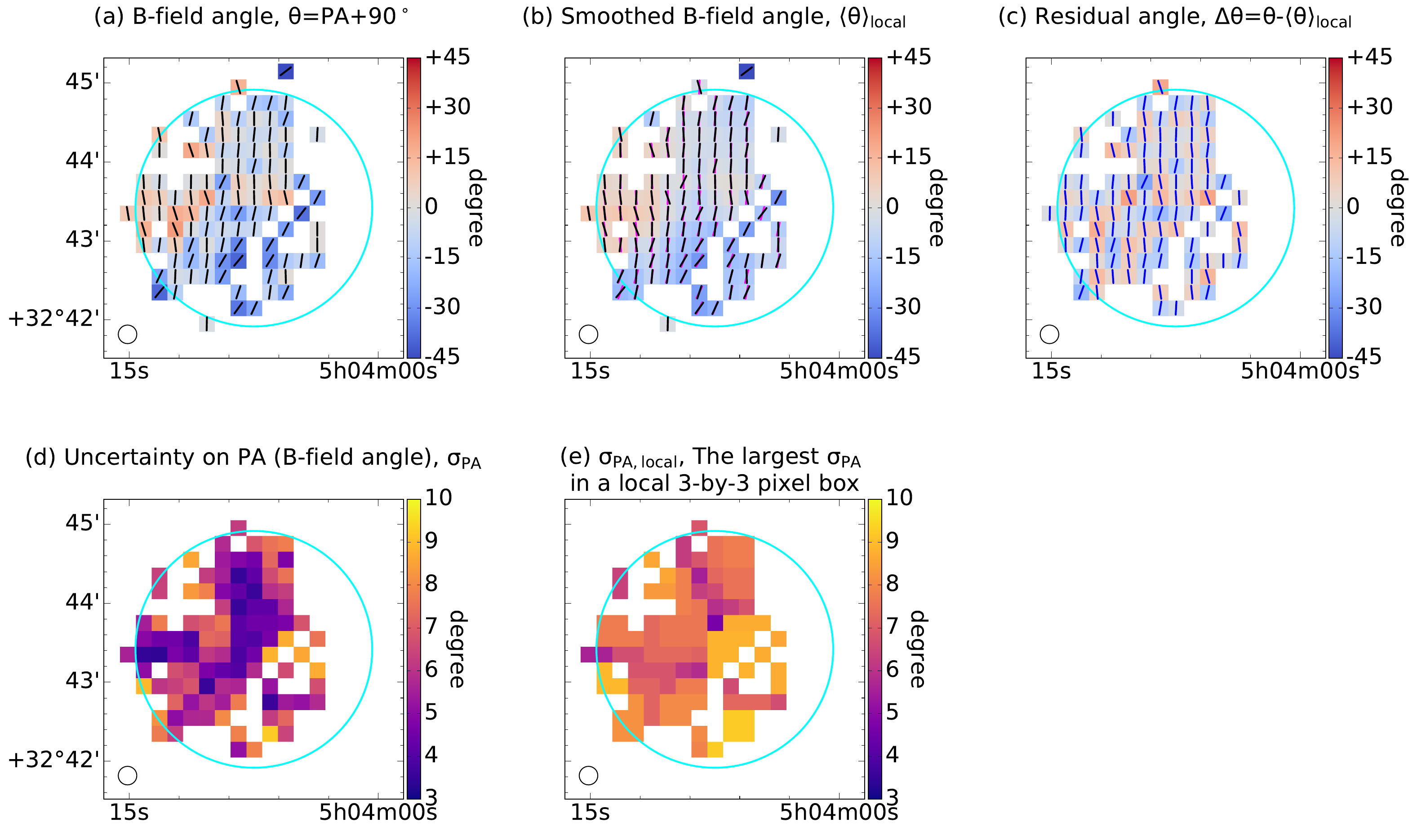}
	\caption{
	Position angles of the \B-field orientations and their uncertainties.
	(a) \B-field position angle map ($\theta$) inferred from the polarization angles (\PA) rotated by 90$^\circ$.
	(b) Mean \B-field angle map ($\langle\theta\rangle_{\rm local}$) from convolving panel (\textit{a}) by a boxcar filter with a size of 3-by-3 pixels.
	The black vectors overlaid on panels (\textit{a}) and (\textit{b}) show the \B-field angles, while the magenta vectors on panel (\textit{b}) show the mean \B-field angles.
	(c) Residual map ($\Delta\theta=\theta-\langle\theta\rangle_{\rm local}$) and blue vectors of these residual angles are overlaid. 
	(d) Uncertainty map ($\sigma_{\PA}$) of \B-field angles. 
	(e) Map of the local largest \PA uncertainty ($\sigma_{\PA, {\rm local}}$), on which each pixel value
    represents the largest PA uncertainty ($\sigma_{\PA}$) among the neighboring nine pixels within a 3-by-3 pixel box centered on that pixel.
    The ``unsharp-masking method'' sets the selection criteria based on $\sigma_{\PA, {\rm local}}$ \citep[see Sec.~\ref{sec:DCF_theta} and also][]{Pattle17}.
    For panels (d) and (e), the minimum, mean, and maximum values are 
    3.5$^\circ$, 5.9$^\circ$, and 9.3$^\circ$ for $\sigma_{\PA}$,
    and 4.7$^\circ$, 7.6$^\circ$, and 9.3$^\circ$ for $\sigma_{\PA, {\rm local}}$, respectively.
	The beam sizes are denoted in the bottom left corners.
	The 3\arcmin POL-2 uniform noise field is indicated by cyan circles.}
    \label{fig:DCF_PA}
\end{figure*}

\begin{figure}
	\centering
    \includegraphics[scale=0.65]{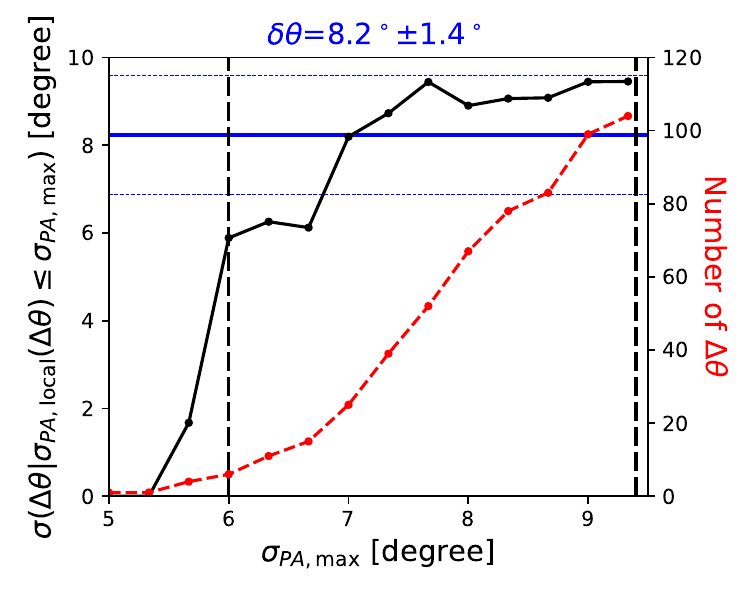}
	\caption{
	Residual angle ($\Delta\theta$) cumulative function of the maximum allowed \B-field angle uncertainty ($\sigma_{\PA, {\rm max}}$), in that the standard deviation of $\Delta\theta$ with the uncertainty ($\sigma_{\PA, {\rm local}}$) equal to or less than $\sigma_{\PA, {\rm max}}$ is evaluated.
	The standard deviation estimate, $\sigma(\Delta \theta)$, is shown as black dots, while the cumulative number of $\Delta\theta$ is shown as red dots.
	The intrinsic \B-field angular dispersion ($\delta \theta$) of 8.2$^\circ$$\pm$1.4$^\circ$ (horizontal solid and dashed blue lines) is determined by computing the mean and standard deviation of the $\sigma(\Delta \theta)$ evaluated from $6^\circ\leq\sigma_{\PA, {\rm max}}\leq9.3^\circ$ (vertical dashed black lines).
    }
    \label{fig:DCF_sig}
\end{figure}

\subsubsection{Magnetic Field Angular Dispersion}\label{sec:DCF_theta}

We use the DCF method to estimate the plane-of-sky magnetic field strength at the core scale with the POL-2 data.
The DCF method requires the intrinsic \B-field angular dispersion measured from the random component of the magnetic field perturbed by local turbulence. However, the \PA measurements include not just the random \B-field component, but also the underlying ordered \B-field component and the \PA measurement uncertainties.
Particularly, the core-scale \B-field shows a slightly curved pattern by changing the \B-field orientation from $\theta_{\rm POL2}\approx0^\circ$ to $\theta_{\rm POL2}\approx-30^\circ$ in the southwestern side (see Fig.~\ref{fig:IQU4}b).
Thus, we adopt the ``unsharp-masking method'' \citep{Pattle17} to estimate the intrinsic angular dispersion of the random \B-field component in L\,1512.
In this method, a smoothed \B-field position angle map is created by convolving the original \B-field angle map with a 3-by-3-pixel boxcar filter (i.e., a width of $\sim$2.6 independent beams). This smoothed map represents the underlying ordered \B-field component.
The 3-by-3-pixel boxcar filter minimizes the influence of the curvature of the field pattern.
After subtracting this smoothed map from the original map, the residual angles represent the random \B-field component and the \PA measurement uncertainties.
The intrinsic \B-field angular dispersion can be estimated by the standard deviation of the residual angles.

Our analysis results for 850~\micron are shown in Figs.~\ref{fig:DCF_PA} and \ref{fig:DCF_sig}.
Figure~\ref{fig:DCF_PA}a shows the \B-field position angle ($\theta$) 12\arcsec-pixel map, where the vectors are shown with uniform lengths for clarity.
Figure~\ref{fig:DCF_PA}d shows the corresponding \B-field-angle uncertainty ($\sigma_{\PA}$) map.
We make a map of the underlying \B-field by smoothing Fig.~\ref{fig:DCF_PA}a with a 3-by-3-pixel boxcar filter (i.e., averaging over the nearby nine pixels).
This smoothed \B-field angle ($\langle\theta\rangle_{\rm local}$) map is shown in Fig.~\ref{fig:DCF_PA}b.
Figure~\ref{fig:DCF_PA}c shows the residual ($\Delta \theta=\theta - \langle\theta\rangle_{\rm local}$) map, in which three isolated pixels from the $\theta$ and $\langle\theta\rangle_{\rm local}$ maps are rejected and 104 pixels remained.

On the residual angle map ($\Delta \theta$), the underlying ordered \B-field geometry has been removed.
In order to estimate a representative intrinsic \B-field angular dispersion ($\delta \theta$) across L\,1512, we present two approaches.
One is the conventional approach, and the other one is a subsequent approach introduced in the unsharp-masking method.
The conventional approach for estimating $\delta \theta$ is via the inverse-variance-weighted quadratic mean of the residual angles, 
\begin{equation}
\delta \theta=\sqrt{\frac{\sum_i{w_i \Delta \theta_i^2}}
{\sum_i{w_i}}},
\end{equation}
where $w_i=1/\sigma_{\PA,i}^2$ 
\citep[e.g.,][]{Clemens16, WangJW19}.
Hence the well-characterized residual angles with smaller $\sigma_{\PA}$ dominate the estimation.
We find $\delta \theta=8.8^\circ\pm 0.5^\circ$ using this approach, where the uncertainty is calculated with the standard error propagation.
On the other hand, \citet{Pattle17} ran Monte Carlo simulations and demonstrated that a representative value of $\delta \theta$ can be better recovered, closer to the input $\delta \theta$ value in their simulations (their so-called ``true value"), via averaging a set of standard deviation estimates of the residual angles selected with different maximum allowed \PA uncertainties ($\sigma_{\PA, {\rm max}}$); i.e., $\langle \sigma(\Delta \theta|{\rm uncertainties} \leq \sigma_{\PA, {\rm max}})\rangle $.
Figure~\ref{fig:DCF_sig} shows these standard deviation estimates as the function of $\sigma_{\PA, {\rm max}}$, with black dots, and the cumulative number of the residual angles limited by $\sigma_{\PA, {\rm max}}$, with red dots.
Figure~\ref{fig:DCF_PA}e shows the angular uncertainty of $\Delta \theta$, denoted by $\sigma_{\PA, {\rm local}}$, adopted by \citeauthor{Pattle17} for this step.
At each pixel, the $\sigma_{\PA, {\rm local}}$ is the largest $\sigma_{\PA}$ of the neighboring nine pixels previously selected to compute $\langle\theta\rangle_{\rm local}$.
\citet{Pattle17} ran a set of Monte Carlo simulations by setting different ``true values" of the intrinsic \B-field angular dispersion ($\delta \theta$) to generate their simulated datasets.
They found the above procedure can approach the input $\delta\theta$ when $\sigma_{\PA, {\rm max}}$ is small and the cumulative number of $\Delta\theta$ is still sufficient. 
They took the mean of such well-characterized standard deviation estimates as the representative estimation of $\delta \theta$ for the whole region.
In our analysis, we collect the standard deviation estimates computed with $6^\circ\leq\sigma_{\PA, {\rm max}}\leq9.3^\circ$ (two vertical dashed black lines in Fig.~\ref{fig:DCF_sig}) because these $\sigma(\Delta \theta)$ estimates remain relatively similar values in contrast to the sharp $\sigma(\Delta \theta)$ decline at $\sigma_{\PA, {\rm max}}<6^\circ$, which is due to the sample of the residual angles being too small (fewer than 6 vectors for this case).
By taking the mean and standard deviation of these $\sigma(\Delta\theta)$ estimates, we obtain an intrinsic \B-field angular dispersion
($\delta \theta$) of ${8.2^\circ\pm1.4^\circ}$ (horizontal blue lines in Fig.~\ref{fig:DCF_sig}) for the whole region.
Compared to the conventional approach, both $\delta \theta$ estimations are consistent, and we adopt $\delta \theta=8.2^\circ\pm1.4^\circ$ for the following analysis.

\begin{figure}
	\centering
    \includegraphics[scale=0.35]{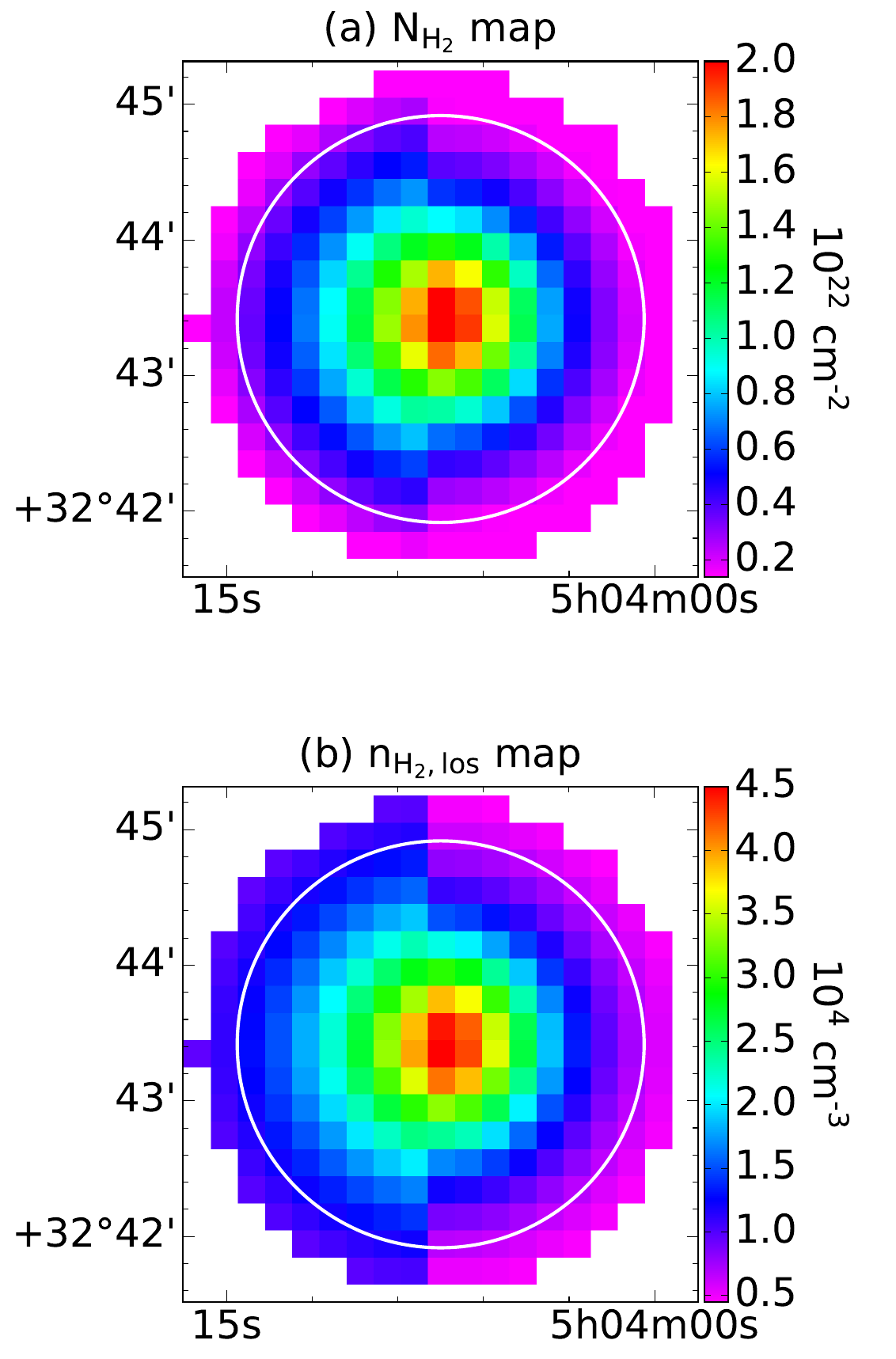}
	\caption{Density maps for the L\,1512 core with $R_{\rm edge}$ = 108\arcsec.
	(a) $N_{\rm H_2}$ map.
    (b) Line-of-sight-averaged $n_{\rm H_2}$ map,
    on which each pixel is calculated as the column density divided by sightline depth.
	The 3\arcmin POL-2 uniform noise field is indicated by white circles for reference.}
    \label{fig:DCF_den}
\end{figure}

\subsubsection{Number Density and Velocity Dispersion}

\citet{Lin20} built an onion-like volume density model of the L\,1512 core, which is comprised of different eastern and western hemispheres. 
They modeled the density structure with Plummer-like profiles of $n_0/(1+(r/R_0)^\eta)$, where $n_0$ is the central density, $R_0$ is the characteristic radius, and $\eta$ is the power-law index of the density profile at $r \gg R_0$.
The best-fit Plummer parameters fitted to \NtHp spectral line observations and dust extinction measurements were found to be 
$n_0=1.1\times10^5$~cm$^{-3}$, $R_0=0.022$~pc, and $\eta=2.0$ for the east side, and
$n_0=1.1\times10^5$~cm$^{-3}$, $R_0=0.027$~pc, and $\eta=3.1$ for the west side.
Since the density required by the DCF method should be the average density along the line of sight, we assume the polarized light is mainly contributed by the L\,1512 core, which has an $R_{\rm edge}$ of 108\arcsec \citep[= 0.073~pc;][]{Lin20}.
With the above density profiles, we 
(1) integrate the volume density profiles within the sphere defined by $R_{\rm edge}$ to obtain a column density ($N_{\rm H_2}$) map, and then
(2) divide the column densities by the corresponding sightline depth inside the sphere to obtain the line-of-sight-averaged H$_2$ number density ($n_{\rm H_2, los}$) map (i.e., for a pixel indexed $j$, $n_{{\rm H_2, los}, j}=N_{{\rm H_2},j}/(2\sqrt{R_{\rm edge, cm}^2-d_{j,{\rm cm}}^2})$, where $R_{\rm edge, cm}$ and $d_{j,{\rm cm}}$ are in the units of centimeter, and $d_{j,{\rm cm}}$ is the projected distance to the core center).
Figure~\ref{fig:DCF_den} shows the $N_{\rm H_2}$ map and $n_{\rm H_2, los}$ map.
For self-consistency, we compute the mean, and standard deviation, of $n_{\rm H_2, los}$ and $N_{\rm H_2}$ values on the same set of pixels, of which the residual angles (Fig.~\ref{fig:DCF_PA}c) are used to estimate the representative \B-field angular dispersion $\delta \theta$.
We find 
$\langle n_{\rm H_2}\rangle=\langle n_{\rm H_2, los}\rangle=(2.1\pm0.9)\times 10^4$~cm$^{-3}$, and 
$\langle N_{\rm H_2}\rangle=(8.3\pm4.7)\times 10^{21}$~cm$^{-2}$.
\citet{Lin20} found that a non-thermal FWHM linewidth of $\Delta v_{\rm NT}=0.109$~km~s$^{-1}$ can reproduce their \NtHp (1--0) spectral observations.
We take their \NtHp (1--0) spectral resolution of 0.031~km~s$^{-1}$ as the uncertainty.
We adopt these values for the DCF analysis and list them in Table \ref{tab:DCF}.

\begin{table}
    \caption{Estimated properties in the DCF analysis from the 850~\micron polarimetry.}
    \begin{tabular}{lr}
    \hline\hline
    Property & Value\\
    \hline
    $\delta \theta$ (degree) & $8.2^\circ\pm1.4^\circ$\\
    $\Delta v_{\rm NT}$ (km s$^{-1}$) & $0.109\pm0.031$\\ 
    $\langle n_{\rm H_2}\rangle$ (cm$^{-3}$) & $(2.1\pm0.9)\times10^4$\\
    $\langle N_{\rm H_2}\rangle$ (cm$^{-2}$) & $(8.3\pm4.7)\times10^{21}$\\
    $B_{\rm pos}$ ($\mu$G) &  $18\pm7$\\
    $\lambda_{\rm obs}$  & $3.5\pm2.4$\\
    $|\textbf{B}_{\rm cor}|$  ($\mu$G)  & $23\pm9$\\
    $\lambda_{\rm cor}$  & $1.2\pm0.8$\\
    $B_{\rm tot}$ ($\mu$G) &  $\sim$32\\
    $B_{\rm los}$ ($\mu$G) &  $\sim$27\\
    $\lambda_{\rm tot}$  & $\sim$1.6\\
    \hline
    \end{tabular}
\tablecomments{Quantities shown are 
the dispersion in magnetic field position angles ($\delta \theta$),
the non-thermal FWHM linewidth ($\Delta v_{\rm NT}$),
the mean H$_2$ number density ($\langle n_{\rm H_2}\rangle$),
the mean H$_2$ column density ($\langle N_{\rm H_2}\rangle$),
the plane-of-sky \B-field strength ($B_{\rm pos}$),
the observed mass-to-flux ratio ($\lambda_{\rm obs}$), 
the statistically corrected total \B-field strength ($|\textbf{B}_{\rm cor}|$),
the statistically corrected mass-to-flux ratio ($\lambda_{\rm cor}$), and 
the total/line-of-sight \B-field strength ($B_{\rm tot}$, $B_{\rm los}$) and the corresponding mass-to-flux ratio ($\lambda_{\rm tot}$) are derived by assuming the L\,1512 core is virially stable (see Sec.~\ref{sec:discussion_sub1}).
The uncertainties are their measured dispersion or computed by the standard error propagation. For $\Delta v_{\rm NT}$, we adopt the \NtHp (1--0) spectral resolution from \citet{Lin20} as the uncertainty.
For $B_{\rm tot}$, $B_{\rm los}$, and $\lambda_{\rm tot}$, we do not estimate uncertainties because their derivation involves energy budgets (\autoref{tab:virial}) that 
may be accurate only to order of magnitude.}
\label{tab:DCF}
\end{table}

\subsubsection{Magnetic Field Strength and Mass-to-flux Ratio}

Using the DCF method (Equation~\ref{equ:DCF}), and the above-estimated values (see Table \ref{tab:DCF}), we calculate the mean plane-of-sky \B-field strength ($B_{\rm pos}$) across the core to be $18\pm7$~$\mu$G, where the uncertainty is computed with the standard error propagation from the dispersions of $\delta \theta$, $\Delta v_{\rm NT}$, and $n_{\rm H_2}$ and hence the uncertainty of $B_{\rm pos}$ represents the dispersion of its distribution.

It is also important to understand if the magnetic field could support the core against gravity, which can be determined by the mass-to-flux ratio in units of the critical ratio, $(M/\Phi)_{\rm crit}$. 
We use the formula from \citet{Crutcher04},
\begin{equation}
\lambda=\frac{(M/\Phi)_{\rm obs}}{(M/\Phi)_{\rm crit}}=7.6\times10^{-21}\frac{N_{\rm H_2}/{\rm cm}^{-2}}{B/{\rm \mu G}}.\label{equ:m2f}
\end{equation}
The core is magnetically supercritical if $\lambda>1$ (i.e., unstable to collapse) and magnetically subcritical if $\lambda<1$ (i.e., magnetically supported).
With our derived $\langle N_{\rm H_2} \rangle$ and $B_{\rm pos}$, we calculate the observed mass-to-flux ratio ($\lambda_{\rm obs}$) as $3.5\pm2.4$ or a range of 1.1--5.9, suggesting that the magnetic field alone may not fully support the core against gravity.
However, the mass-to-flux ratio could be overestimated by $\lambda_{\rm obs}$, due to the unknown inclination of the \B-field and the 3D geometry of the core \citep{Crutcher04}. 
On the other hand, the $\lambda_{\rm obs}$ lower limit of 1.1 does not entirely rule out the possibility of the magnetically critical condition, owing to various sources of uncertainty in the DCF method \citep{Crutcher12, Pattle19FrASS}.
Therefore, our estimation of the mass-to-flux ratio suggests that the L\,1512 core is approximately magnetically critical or supercritical.
We note that the mass-to-flux ratio calculated here is considered as a qualitative indicator rather than as a precise measure of the L\,1512 core's stability against gravity.
We also note that when computing the mass-to-flux ratio, contributions from thermal and non-thermal energies are not considered.

\citet{Crutcher04} showed that $B_{\rm pos}$ and $\lambda_{\rm obs}$ can be statistically corrected by averaging the cases of all \B-field inclinations with respect to the line of sight.
Under this methodology, $B_{\rm pos}$ is a statistical average of the total magnetic field, \textbf{B}$_{\rm cor}$, which can point over a total solid angle of 2$\pi$ subtended from the core center.
The correction is
\begin{equation}
B_{\rm pos}=\frac{\pi}{4}|\textbf{B}_{\rm cor}|. \label{equ:B_3D}
\end{equation}
In this case, the total magnetic field would have a strength of $23\pm9$~$\mu$G.
For our observed mass-to-flux ratio, $\lambda_{\rm obs}$, \citet{Crutcher04} derived the correction as
\begin{equation}
\lambda_{\rm cor}=\frac{\lambda_{\rm obs}}{3}.
\end{equation}
This correction also takes into account the projection effect of $N_{\rm H_2}$, assuming that the magnetic flux tube is aligned with the core minor axis.
This yields a mass-to-flux ratio of $\lambda_{\rm cor}$ = $1.2\pm0.8$, still suggesting the core is approximately magnetically critical or slightly supercritical. 
We will further discuss the \B-field strength in the context of the Virial analysis in Sec.~\ref{sec:discussion_L1512_Bf}.

\subsection{Grain Alignment}

\begin{figure}
    \includegraphics[scale=0.5]{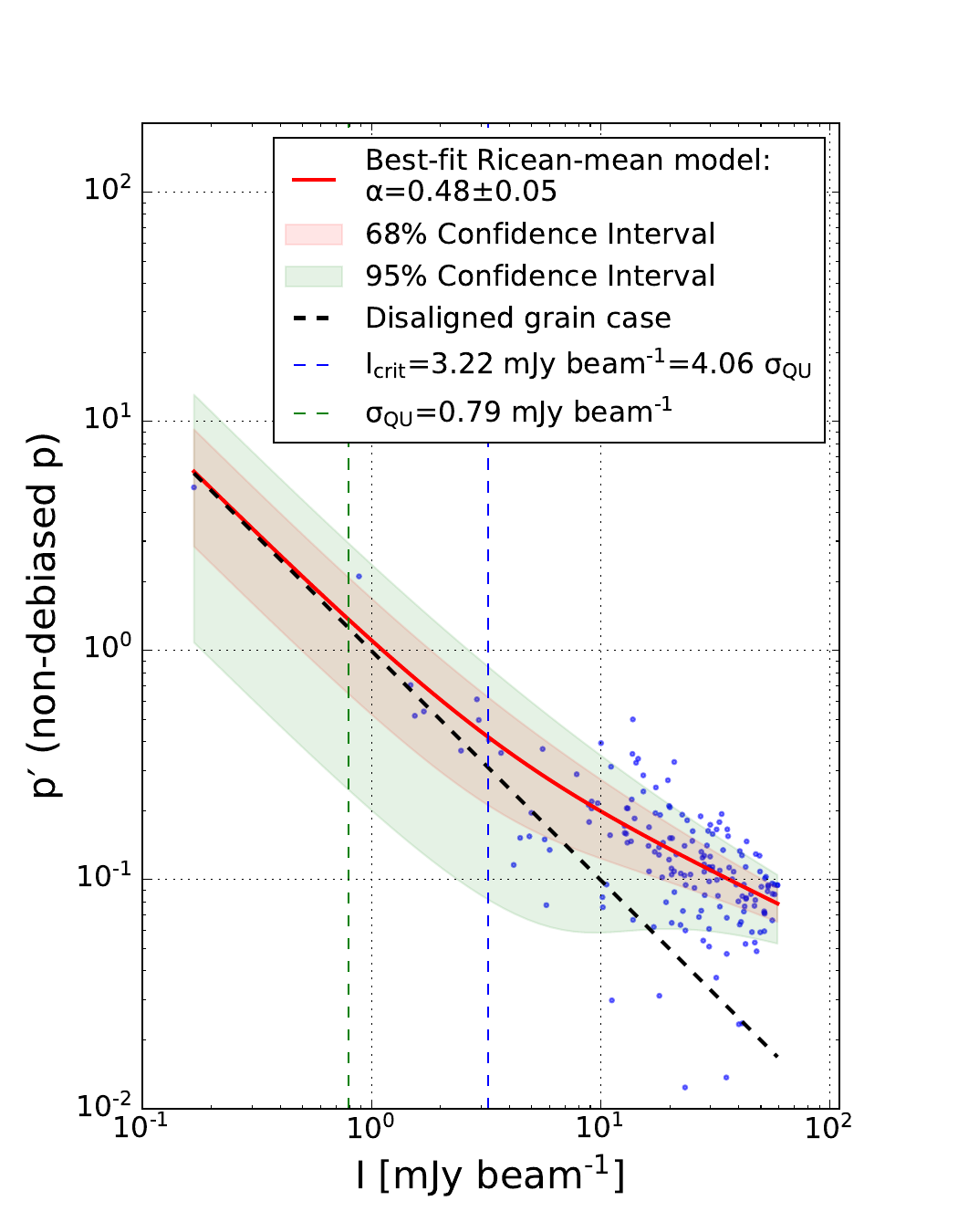}
    \centering
	\caption{
	JCMT 850~\micron non-debiased polarization fraction ($p\arcmin$) is plotted against the total intensity (Stokes \I).
    The best-fit Ricean-mean model (Equation~\ref{equ:p_align}) and the grain disalignment model (Equation~\ref{equ:p_unalign}) are shown as solid red and dashed black lines, respectively. 
    The red- and green-colored regions are the 68\% and 95\% confidence intervals, respectively.}
    \label{fig:p_eff}
\end{figure}

A long-standing debate about submillimeter polarimetry is whether dust grains remain aligned with magnetic fields inside starless cores \citep[e.g.,][]{Alves14, Jones15, Andersson15, WangJW19, Pattle19}.
From a theoretical standpoint, the main alignment mechanism is thought to be Radiative Alignment Torques (\B-RATs), in which an anisotropic radiation field is a key to aligning the dust grains with the magnetic field \citep[e.g.,][]{Dolginov76, Lazarian07, Andersson15}.
In the center of starless cores, the absence of an internal radiation source producing an anisotropic radiation field might lead to an outcome where dust grains are not aligned with the magnetic field.
Therefore, it is crucial to investigate whether the dust grains are aligned at the centers of starless cores.

To evaluate grain alignment in the submillimeter regime, a power-law index in the $p-I$ relation\footnote{We note that the polarization fraction discussed in this power law is the true/intrinsic value ($p_{\rm true}$), while both of the non-debiased and debiased polarization fractions ($p\arcmin$ and $p$ defined by Equations \ref{equ:p_prime} and \ref{equ:p}) are measurements. The measurement $p$ is an attempt made by the asymptotic estimator to find $p_{\rm true}$.},
\begin{equation}
p_{\rm true}(I)\propto I^{-\alpha},
\end{equation}
is sought,
where $0\leq \alpha \leq1$ \citep[e.g., ][]{Jones15, Alves15}. 
In the $\alpha=0$ case, the polarization fraction is constant at every depth along each sightline, indicating a perfectly aligned case.
For the $\alpha=1$ case, the polarized intensity ($=p_{\rm true}\cdot I$) is constant.
In such a case, one interpretation is that only the outer shell of the starless core contributes to the polarized intensity, and no contribution comes from the inner region, implying a completely disaligned case \citep{Pattle19}. 
We can only probe the \B-field morphology at the core surface in this case. 

In order to determine the index $\alpha$, we should take into account that the polarization fraction measurements do not follow a Gaussian distribution.
Typically, the polarization fraction measurements are statistically debiased with a polarization estimator \citep[e.g., the asymptotic estimator defined by Equation~\ref{equ:p}; please refer to][for the detailed discussion of other estimators]{Montier15} and filtered with an SNR criterion for fitting $\alpha$. 
In fact, these polarization estimators are \textit{biased} estimators but they work better in the high-SNR regime because the \textit{bias} is minimized \citep{Montier15}; therefore, the low-SNR data should be removed in order to better determine $\alpha$.
For faint starless cores, such SNR criteria may remove too much data, and thus $\alpha$ might not be well-constrained.

Instead of debiasing data and removing the low-SNR ones, \citet{Pattle19} assumed that the non-biased polarization fraction measurements follow a Rice distribution, including both low- and high-SNR data, and took the mean of the Rice distribution \citep{Rice45, Serkowski58} as the estimator of the true polarization fraction at a given \I. 
They referred to this estimator as the Ricean-mean model,
\begin{equation}
p\prime(I) = \sqrt{\frac{\pi}{2}}\left(\frac{I}{\sigma_{QU}}\right)^{-1}
\mathcal{L}_{\frac{1}{2}}\left(-\frac{p_{\sigma_{QU}}^2}{2}\left(\frac{I}{\sigma_{QU}}\right)^{2(1-\alpha)}\right),
\label{equ:p_align}
\end{equation}
where $\mathcal{L}_{\frac{1}{2}}$ is a Laguerre polynomial of order 1/2, 
$\sigma_{QU}$ is the average of $\sigma_{Q}$ and $\sigma_{U}$,
$\alpha$ is the power-law index, and 
$p_{\sigma_{QU}}$ is the expected polarization fraction when $I=\sigma_{QU}$.
We note that technically, $p_{\sigma_{QU}}$ and $\sigma_{QU}$ are the free scaling factors of the assumed underlying $p-I$ relation of
$p_{\rm true}=p_{\sigma_{QU}}(I/\sigma_{QU})^{-\alpha}$,
where the \I scaling factor is chosen to be the practical noise ($\sigma_{QU}$) while the polarization fraction scaling factor ($p_{\sigma_{QU}}$) is left for fitting.
This Ricean estimator allows for a better estimation of the index $\alpha$ by being able to properly include the low-SNR data.

We use our data to determine the index $\alpha$ in L\,1512 with the above Ricean-mean model.
We collect data within a 3\arcmin uniform noise field ({${\sigma_{QU}=0.79}$~\mJybm}), and supply $\sigma_p$ as a data weighting in order to perform a fitting with the Python \texttt{scipy.optimize.curve\_fit} function.
We obtain a best-fit $\alpha$ of ${0.48\pm0.05}$, and $p_{\sigma_{QU}}$ of ${0.60\pm0.10}$.
Figure~\ref{fig:p_eff} shows the best-fit Ricean-mean model in the solid red curve.
The true value of ${\alpha=0.48\pm0.05}$ is recoverable above the critical intensity ($I_{\rm crit}$) of 3.22~\mJybm (corresponding to a SNR of $I_{\rm crit}/\sigma_{QU}=4.05$).
In contrast, owing to the low SNR, the true $\alpha$ is not recoverable below $I_{\rm crit}$ but apparently approaches the value of unity expected for disaligned grains,
\begin{equation}
p\prime(I) = \sqrt{\frac{\pi}{2}}\left(\frac{I}{\sigma_{QU}}\right)^{-1},
\label{equ:p_unalign}
\end{equation}
shown as the dashed black line. 
The index of ${\alpha=0.48\pm0.05}$ suggests that some of the dust grains remain aligned with the magnetic field at higher densities \citep{Andersson15, Pattle19FrASS}.
Therefore, our data are able to probe the magnetic field in the L\,1512 core.

\section{Discussion}\label{sec:discussion_L1512_Bf}

\subsection{Gravitational Stability}\label{sec:discussion_sub1}

The gravitational stability of the L\,1512 core can be assessed using the Virial theorem \citep{Mckee92, McKee07}, which can be written as
\begin{equation}
    \frac{1}{2}\frac{d^2I}{dt^2}=2(E_{\rm thermal}-E_{\rm surface}+E_{\rm turb}+E_{\rm rot})+E_{\rm grav}+E_{\rm mag},\label{equ:vir}
\end{equation}
where $I$ is the moment of inertia, 
$E_{\rm thermal}$ is the thermal energy in the core, 
$E_{\rm surface}$ is the surface kinetic term due to external pressure,
$E_{\rm turb}$ is the turbulent energy, 
$E_{\rm rot}$ is the rotational energy,
$E_{\rm grav}$ is the gravitational potential energy,
and $E_{\rm mag}$ is the magnetic energy
calculated by the total \B-field strength ($B_{\rm tot}$).
The magnetic energy can be divided into the plane-of-sky and line-of-sight components such that $E_{\rm mag}=E_{\rm mag, pos}+E_{\rm mag, los}$ since $B_{\rm tot}^2=B_{\rm pos}^2+B_{\rm los}^2$.
The total kinetic energy can be defined as $T=E_{\rm thermal}-E_{\rm surface}+E_{\rm turb}+E_{\rm rot}$.
By defining the virial energy of $E_{\rm vir}=2T+E_{\rm grav}$, we rewrite Equation~\ref{equ:vir} as
\begin{equation}
    \frac{1}{2}\frac{d^2I}{dt^2}=E_{\rm vir}+E_{\rm mag, pos}+E_{\rm mag, los}.\label{equ:vir2}
\end{equation}
The sign of $d^2I/dt^2$ determines whether the core is virially unbound with a positive net energy ($d^2I/dt^2>0$) or virially bound with a negative net energy ($d^2I/dt^2<0$), and $d^2I/dt^2=0$ means that the core is virially stable.

To assess the core stability, the physical structure of L\,1512 is needed. \citet{Lin20} focused on \NtHp multi-transition spectra modeling and visual extinction measurement to build an onion model to describe the volume density, kinetic temperature, and rotational velocity profiles for the case of a constant turbulent velocity in the L\,1512 core.
Their onion model can well reproduce the \NtHp and the other line data observed along a horizontal (RA) cut and a vertical (Dec) cut across the L\,1512 core.
The onion model is comprised of eastern (nine layers) and western (six layers) hemispheres, because the west extent of L\,1512 is a factor of $\sim$2/3 shorter than along the eastern side.
Here, we adopt the eastern hemisphere model to represent the entire core because the eastern side of L\,1512 is more spherically symmetric with respect to the core center than the western side.
Thus the eastern hemisphere onion model could provide a better description for the core \citep[also see Figs.~1 and 2 from][]{Lin20}.

\begin{table}
    \caption{Energy budgets in L\,1512.}
    \begin{tabular}{lr}
    \hline\hline
    Energy compared with $E_{\rm grav}$ & Value\\
    \hline
    $E_{\rm grav}$ (erg) & $-3.9\times 10^{42}$\\
    $E_{\rm thermal}/|E_{\rm grav}|$ & 0.61\\
    $E_{\rm surface}/|E_{\rm grav}|$ & 0.41\\
    $E_{\rm turb}/|E_{\rm grav}|$ &  0.04\\
    $E_{\rm rot}/|E_{\rm grav}|$ & 0.01\\
    $T/|E_{\rm grav}|$ & 0.25\\
    $E_{\rm vir}/|E_{\rm grav}|$ & $-$0.51\\
    $E_{\rm mag, pos}/|E_{\rm grav}|$ & 0.16\\
    $(E_{\rm vir}+E_{\rm mag, pos})/|E_{\rm grav}|$ & $-$0.35\\
    \hline
    \end{tabular}
\tablecomments{Quantities shown are 
the gravitational energy ($E_{\rm grav}$), 
the thermal energy ($E_{\rm thermal}$), 
the surface kinetic term due to external pressure ($E_{\rm surface}$), 
the turbulent energy ($E_{\rm turb}$), the rotational energy ($E_{\rm rot}$), 
the total kinetic energy ($T=E_{\rm thermal}-E_{\rm surface}+E_{\rm turb}+E_{\rm rot}$),
the virial energy ($E_{\rm vir}=2T+E_{\rm grav}$), 
and the plane-of-sky component magnetic energy ($E_{\rm mag, pos}$).}
\label{tab:virial}
\end{table}

We use the physical parameters in the eastern onion model from \citet{Lin20} to calculate each energy term in the Virial equation (Equation~\ref{equ:vir}).
Please refer to Table C.1 and Fig.~5c from \citet{Lin20} for the density, temperature, and rotational velocity profiles.
In addition, the constant one-dimensional non-thermal velocity dispersion, $\sigma_{v, {\rm NT}}$, used in the onion model is 0.046~km~s$^{-1}$.
We calculate the plane-of-sky component magnetic energy ($E_{\rm mag,pos}$) by using the DCF-derived plane-of-sky magnetic field strength ($B_{\rm pos}$) of 18~$\mu$G (\autoref{tab:DCF}).
\autoref{tab:virial} summarizes the results for each energy of the L\,1512 core and
\autoref{app:virial} shows the formulae we used for deriving each energy term.
If we use the physical parameters in the western onion model from \citet{Lin20}, the virial energy ($E_{\rm vir}$) will change from $-$0.51 to $-$0.39 and 
the plane-of-sky component magnetic energy ($E_{\rm mag, pos}$) will change from 0.16 to 0.11, in units of $|E_{\rm grav}|$.
Accordingly, if $E_{\rm mag,los}=0$, the value of the inertia term on the left-hand side of Equation~\ref{equ:vir2} will change from $-$0.35 to $-$0.28 in units of $|E_{\rm grav}|$.
We note that these energies could also have significant uncertainties; thus, these derived values are accurate only to order of magnitude.
However, as the dominant uncertainty is mass, all the derived energies are linearly dependent on the mass, except that $E_{\rm grav}$ is dependent on the square of mass.

The rotational energy ($E_{\rm rot}$) in \autoref{tab:virial} is one of the interpretations from the \NtHp (1--0) observation toward L\,1512, which reveals a uniform velocity gradient (2.26$\pm$0.04~km~s$^{-1}$~pc$^{-1}$) along roughly the north-south direction across an extent of $\sim0.1$~pc \citep{Caselli02, Lin20}. 
An interpretation is a slow solid-body rotation (\citealt{Caselli02}; Fig.~5c in \citealt{Lin20}), of which the corresponding rotation period of 2.72$\pm$0.05~Myr ($\Omega=(7.3\pm0.1)\times10^{-14}$~rad~s$^{-1}$ and $v_{\rm rot}\leq0.07$~km~s$^{-1}$) is about twice as long as the core lifetime of $\gtrsim$1.4~Myr \citep[][]{Lin20}. 
The corresponding $E_{\rm rot}$ is only 1\% of $|E_{\rm grav}|$. 
This contribution does not affect the sign of $d^2I/dt^2$.
Another interpretation involves the projection of tilted inward motions along filaments threaded by magnetic flux tubes \citep{Balsara01}, which can be related to the slow, subparsec-scale accretion flows toward the core along the north-south CO filament \citep{Falgarone01}. 
The blue- and red-shifted inward motions may lead to a kink of magnetic fields that contributes to the deviations of POL-2 polarization vectors from being uniform, as shown in Fig.~\ref{fig:IQU4}b.

The plane-of-sky magnetic energy ($E_{\rm mag, pos}$) in \autoref{tab:virial} is calculated with the DCF-derived plane-of-sky \B-field strength ($B_{\rm pos}$) of 18~$\mu$G. Because ${E_{\rm vir}+E_{\rm mag, pos}=-0.35|E_{\rm grav}|<0}$, the L\,1512 core is virially bound if $E_{\rm mag, los}=0$, and further contraction could happen.
However, the molecular spectral line observations toward L\,1512 do not suggest that L\,1512 is a ``contracting core" but instead suggest it may have an ``oscillating envelope."
Using the high-density tracer \NtHp (1--0), \citet{Lin20} performed non-LTE radiative transfer modeling and found no significant infall in the L\,1512 core region. 
Employing the same radiative transfer model, we estimate an upper limit for the radial infall velocity of $\sim$0.04~km~s$^{-1}$ by analyzing their \NtHp data, indicating that the core region is relatively quiescent. 
\citeauthor{Lin20} also found that fitting their multi-line observations of \NtHp, \NtDp, \DCOp, and o-\HtDp does not require an infall velocity field in the radiative transfer model, even though the hyperfine structures were carefully considered.
Additionally, the envelope of L\,1512 was suggested to be oscillating because a mixture of blue and red asymmetric spectral line profiles was observed across the entire cloud in the CS (2--1) line \citep{Lee99a, Lee01, Lee11}, which is a low-density envelope tracer significantly depleted in the central core region. 
Another envelope tracer, HCN (1--0), which suffers fewer depletion effects compared with CS, also shows a mixed spectral feature across the observing area \citep{Sohn07, Kim16}.
Although \citet{Schnee13} found a red asymmetric feature from their HCO$^+$ (3--2) observation, indicating an outward motion, their single-pointing observation does not conflict with the oscillation notion.
Based on these spectral observations, L\,1512 is likely a long-lived starless core, which is consistent with the core lifetime estimated to be longer than 1.4~Myr according to deuteration chemical modeling \citep{Lin20}.
Therefore, we expect the L\,1512 core to be approximately virially stable.
However, neither the kinetic pressure nor magnetic pressure of the plane-of-sky \B-field alone can support the L\,1512 core (i.e., $2T<|E_{\rm grav}|$ and $E_{\rm mag, pos}<|E_{\rm grav}|$).
Given that the magnetic field seems to thread from the large scale to the core scale in L\,1512, and the plane-of-sky  magnetic field maintains a well-ordered field pattern, we speculate that the magnetic field is not entirely dynamically unimportant in the L\,1512 core.
It can be the case that the magnetic field does not lie close to the sky plane and a hitherto hidden line-to-sight magnetic field provides additional support, making L\,1512 nearly stable.
While Zeeman measurements of L\,1512 would help to estimate the line-to-sight \B-field strength, such a measurement is currently unavailable.

If the magnetic field is strong enough for the total magnetic energy to compensate for the negative $E_{\rm vir}$ value, leading the L\,1512 core to be virially stable
(i.e., ${0=\frac{1}{2}\frac{d^2I}{dt^2}=E_{\rm vir}+E_{\rm mag}}$ implies ${E_{\rm mag}=0.51|E_{\rm grav}|}$; see \autoref{app:virial}),
the total \B-field strength ($B_{\rm tot}$) would need to be $\sim$32~$\mu$G. 
This strength is within the range measured in other starless cores derived via the DCF technique \citep[e.g.,][]{Kirk06, Pattle21, Myers21} and via the Zeeman effect technique \citep[e.g.,][]{Crutcher00, Troland08}.
In this case, the line-of-sight \B-field component ($B_{\rm los}=\sqrt{B_{\rm tot}^2-B_{\rm pos}^2}$) is estimated to be $\sim$27~$\mu$G and the inclination angle ($i=\sin^{-1}(B_{\rm pos}/B_{\rm tot})$) of total \B-field direction is $\sim$34$^\circ$ with respect to the line of sight. 
We note that the derived values of $B_{\rm tot}$ and $B_{\rm los}$, as well as the DCF-derived $B_{\rm pos}$ strength and energy budgets, carry certain uncertainties.
It is better to consider that if $B_{\rm los}$ is present with a similar order of magnitude to $B_{\rm pos}=18\pm7$~$\mu$G, these values suggest an approximate virial stability of L\,1512.
By adopting $N_{\rm H_2}=\langle N_{\rm H_2} \rangle \cos(i)$ and $B=B_{\rm tot}$ in Equation~\ref{equ:m2f}, the corresponding mass-to-flux ratio ($\lambda_{\rm tot}$) is $\sim$1.6, suggesting an approximately magnetically critical or sightly supercritical condition.
Although $B_{\rm tot}$ does not make the L\,1512 core magnetically subcritical, the magnetic pressure and the kinetic pressure are of comparable importance ($2T\sim E_{\rm mag}$ and $2T + E_{\rm mag} \sim |E_{\rm grav}|$) in supporting the core.
On the other hand, if the magnetic fields in the L\,1512 core do not have a line-of-sight component or just have a small $B_{\rm los}$, L\,1512 would be magnetically supercritical and should be collapsing. 
However, the aforementioned spectral observations show this is not the case.
Therefore, either L\,1512 has just recently reached supercriticality and will collapse at any time and we happened to observe it in this special state,
or L\,1512 is nearly stable and there is an as-yet unseen line-of-sight \B-field.

\subsection{Relationship between Large- to Core-scale Magnetic Fields}

The \H band polarimetry is an important tool to trace the magnetic field at scales between those observed with POL-2 and with Planck. 
Our POL-2 850~\micron, Mimir \H band, and the Planck polarization observations enable efficient plane-of-sky \B-field characterization across the small, intermediate, and large scales of the L\,1512 cloud.
As shown in Figs.\,\ref{fig:IQU4} and \ref{fig:NIR}, the magnetic field orientation of the L\,1512 envelope (with an average field angle of $\theta_{\rm H}=-{15^\circ}_{-39^\circ}^{+40^\circ}$) appears to be inherited from that of the large-scale \B-field ($\theta_{\rm Planck}=-30^\circ$) and reveals a twist in the \B-field morphology between the core and the envelope. 
The twist occurs in the southwestern core region, where the field angle $\theta_{\rm POL2}$ bends from $\approx$$0^\circ$ in the core center to $\approx$$-30^\circ$, aligning with the nearby envelope-scale field at $\theta_{\rm H}\approx-30^\circ$ (see Sec.~\ref{sec:Bfield}).

While the 3D field geometry of the field lines remains unknown, the observed twisted field may hint that the matter altered the field at the core scale of $\sim$0.1~pc, with the field orientation still close to the initial large-scale field. 
This suggests that L\,1512 is likely in an intermediate phase, transitioning from the magnetically dominated phase (i.e., magnetically subcritical phase) to the matter-dominated phase (i.e., magnetically supercritical phase).
This intermediate phase was proposed by \citet{Ward-Thompson23} based on their polarimetric observations of nine starless cores embedded within the L\,1495A-B10 filaments. The authors found that the plane-of-sky core-scale \B-field orientations of these cores are roughly perpendicular to the filaments. However, they are not correlated with the large-scale \B-field orientations measured by Planck, except for the lowest-density and possibly youngest core, where the core-scale field is still close to the large-scale field. 
In this case, 
a twisted field may be due to the early mass accumulation in the core, and the local field would become perpendicular to the major axis of the core if the gravitational instability is further enhanced.

On the other hand, twisted magnetic fields are also observed in protostellar sources such as filamentary gas flows on the subparsec scale in Serpens South \citep{Pillai20} and a twist of $\sim$45$^\circ$ within the inner $\sim$0.005~pc region of L\,483, a Class 0 protobinary core \citep{Cox22}.
These authors suggested that these twists result from the gas flow feeding onto the nearby cluster-forming regions and interactions involving binary systems.
Thus, the aforementioned protostellar twists are 
formed in the magnetically supercritical phase, 
where gravity can efficiently influence the local magnetic field orientations.
In contrast, the twist observed in L\,1512 is likely developed during the intermediate phase.

In terms of kinematics, the L\,1512 core exhibits quiescent gas motions.
With the lack of significant gravitational instability, the magnetic field may be dynamically important in the core evolution of L\,1512.
Moreover, since the L\,1512 cloud was reported to be magnetically subcritical based on \R band data \citep[$\lambda_{\rm obs}\sim0.8$ at the scale of $\sim$0.8~pc;][]{Sharma22}, the L\,1512 core ($\lambda_{\rm obs}=3.5\pm2.4$ and $\lambda_{\rm tot}\sim1.6$ at the scale of $\sim$0.1~pc) may have undergone a sub-to-supercritical transition, through such as ambipolar diffusion.

As mentioned in Sec.~\ref{sec:intro}, only a few starless cores have resolved submm polarization detections.
Among these cores, L\,183 \citep{Clemens12a, Karoly20},  FeSt\,1-457 \citep{Alves14, Kandori17a}, and L\,1544 \citep{Clemens16} also had polarization detection in the \H band.
In the cases of L\,1544 and FeSt\,1-457, despite the presence of some non-uniform \B-field structures \citep{Clemens16, Alves14}, the orientations of their core-, envelope-, and large-scale magnetic fields remain roughly consistent, similar to L\,1512.
In contrast, L\,183 is the only core in this sample where the orientation of the core-scale \B-field tends to be perpendicular to that of the large-scale \B-field; 
the transition of the \B-field in the L\,183 from its envelope to the core was captured by the \H band polarization measurements \citep{Clemens12a, Karoly20}.
The apparently similar orientations in the large- and core-scale magnetic fields in L\,1512, L\,1544, and FeSt\,1-457 (unless due to a projection effect) suggests that material can accrete onto these cores along magnetic field lines more efficiently than in the case in L\,183.
Such core formation scenario has been demonstrated by \citet{ChenCY20} with their turbulent MHD simulations, suggesting that dense cores could accumulate more mass when the core-scale \B-fields are aligned with the parsec-scale \B-field. 
However, the L\,183 cloud has a total mass of $\sim$80 M$_{\odot}$ \citep{Pagani04}, which is considerably larger than the total mass of the L\,1544 cloud, estimated to be $\sim$10 M$_{\odot}$ \citep{KimSY22}.
It could be possible that the L\,183 cloud is older, enabling it to accumulate sufficient mass, or that the L\,183 cloud was supplied with a substantial ancient mass reservoir.

The potential transition from subcritical to supercritical magnetic conditions through ambipolar diffusion is a shared characteristic among L\,1512, L\,1544, and FeSt\,1-457. 
This phenomenon has been proposed specifically for L\,1544 \citep{Ciolek00, Li02} and FeSt\,1-457 \citep{Kandori20a, Bino21}. 
In terms of kinematics, L\,1544 exhibits extended inward motions \citep{Tafalla98}, whereas L\,1512 and FeSt\,1-457 have quiescent cores and oscillating envelopes \citep{Aguti07, Lee11, Juarez17, Lin20}. 
Moreover, FeSt\,1-457 is suggested to be supported by both kinetic pressure and magnetic pressure \citep{Kandori18b}, similar to L\,1512.
In contrast, the entire L\,183 core is found to be subcritical according to POL-2 observations \citep{Karoly20}, suggesting the dominance of ambipolar diffusion in its core evolution. 
This could be consistent with the absence of inward motions for the L\,183 core \citep{Pagani07} and that its surrounding envelope is suggested to be oscillating \citep{Schnee13}. 
Despite that, the L\,183 core has developed a central density \citep[$2.3\times10^6$ cm$^{-3}$;][]{Pagani07} comparable to that of L\,1544 \citep[$8.6\times10^6$ cm$^{-3}$;][]{Keto15, Sipila22}.
Further observations of more starless cores remain vital to shed light on their core formation process and to better understand their environmental influences.

\section{Conclusions}\label{sec:conclusion_L1512_Bf}

We present JCMT POL-2 850~\micron dust continuum polarization observations and 
Mimir \H-band NIR polarization observations toward L\,1512. 
Our observations reveal an ordered core-scale \B-field morphology in L\,1512.
From our analysis, we find the following:

\begin{enumerate}
\item The L\,1512  850~\micron data, as obtained, likely suffer from missing large-scale flux for the POL-2 data collection. 
We found that principal component analysis (PCA) in the standard reduction process removed extended emission, resulting in apparent non-detection of the total intensity.
By including a SCUBA-2 Stokes \I map in the reduction procedure,
POL-2 Stokes \Q and \U maps could be correctly recovered.

\item The magnetic field traced by POL-2 850~\micron, Mimir \H band, AIMPOL \textit{R} band, and Planck polarization data are in agreement as to the average field orientation, suggesting that the large-scale \B-field threads the L\,1512 cloud down into the dense core region.
The largest angular dispersion, found in Mimir \H band data, indicates that a transition of \B-field morphology could be happening at the envelope scale.

\item Ricean-mean modeling of the non-debiased polarization fraction data yielded a power-law index $\alpha$ of 0.48$\pm$0.05 in the $p^\prime\propto I^{-\alpha}$ relation, indicating the dust grains retain substantial alignment with the magnetic field at the higher densities within the core.

\item A Davis–Chandrasekhar–Fermi analysis revealed a plane-of-sky \B-field strength of 18$\pm$7~$\mu$G, and a mass-to-flux ratio of $\lambda_{\rm obs}=3.5\pm2.4$ or a range of 1.1--5.9, suggesting that L\,1512 is magnetically supercritical; 
however, the true mass-to-flux ratio may be being overestimated by $\lambda_{\rm obs}$, and the magnetically critical condition is not entirely ruled out.
Given the absence of significant inward motions, and the presence of a well-ordered core-scale \B-field and an oscillating envelope, it is likely that L\,1512 \textit{\textbf{is}} supported by both magnetic and kinetic pressures.
By assuming L\,1512 is virially stable and including the kinetic energy, we estimated that a total \B-field strength of $\sim$32~$\mu$G could support the L\,1512 core against gravity, suggesting a corresponding mass-to-flux ratio of $\sim$1.6.
This requires a hitherto hidden line-of-sight \B-field component of $\sim$27~$\mu$G, which could be sought using Zeeman effect techniques.

\item Alternatively, if there is little to no line-of-sight \B-field, then L\,1512 should be collapsing. In this case, L\,1512 may have just recently reached supercriticality and will collapse at any time.

\end{enumerate}

\acknowledgments
We thank the referee for providing thoughtful and
constructive feedback that helped to improve this article.
We also thank Tyler Bourke for his comments on this manuscript that helped to improve it.
S.J.L. and S.P.L. acknowledge the support from the Ministry of Science 
and Technology (MOST) of Taiwan with grant MOST 106-2119-M-007-021-MY3 
and MOST 109-2112-M-007-010-MY3.
This work used high-performance computing facilities operated by the
Center for Informatics and Computation in Astronomy (CICA) at National
Tsing Hua University. This equipment was funded by the Ministry of
Education of Taiwan, the Ministry of Science and Technology of Taiwan,
and National Tsing Hua University.
Nawfel Bouflous and Patrick Hudelot (TERAPIX data center, IAP, Paris, France) 
are warmly thanked for their help in preparing the CFHT/WIRCAM observation 
scenario and for performing the data reduction.
This work was supported by the Programme National “Physique et Chimie du 
Milieu Interstellaire” (PCMI) of CNRS/INSU with INC/INP co-funded by CEA 
and CNES and by Action F\'ed\'eratrice Astrochimie de l'Observatoire de Paris.
The James Clerk Maxwell Telescope is operated by the East Asian Observatory on behalf of The National Astronomical Observatory of Japan; Academia Sinica Institute of Astronomy and Astrophysics; the Korea Astronomy and Space Science Institute; the National Astronomical Research Institute of Thailand; Center for Astronomical Mega-Science (as well as the National Key R\&D Program of China with No. 2017YFA0402700). Additional funding support is provided by the Science and Technology Facilities Council of the United Kingdom and participating universities and organizations in the United Kingdom and Canada.
Additional funds for the construction of SCUBA-2 were provided by the Canada Foundation for Innovation. 
The authors wish to recognize and acknowledge the very significant cultural role and reverence that the summit of Maunakea has always had within the indigenous Hawaiian community.  We are most fortunate to have the opportunity to conduct observations from this mountain.
This research was conducted in part using the Mimir instrument, jointly developed at Boston University and Lowell Observatory and supported by NASA, NSF, and the W.M. Keck Foundation.
Analysis software for Mimir data was developed under NSF grants AST 06-07500 and 09-07790 to Boston University.
This research used the facilities of the Canadian Astronomy Data Centre operated by the National Research Council of Canada with the support of the Canadian Space Agency.
This research has made use of the NASA/IPAC Infrared Science Archive, which is 
operated by the Jet Propulsion Laboratory, California Institute of Technology, 
under contract with the National Aeronautics and Space Administration.

\facilities{James Clerk Maxwell Telescope (JCMT), Perkins Telescope Observatory.}
\software{
Starlink \citep{Currie14}, 
Mimir Software Package \citep{Clemens12b},
APLpy \citep{aplpy2012, aplpy2019},
Astropy \citep{astropy1, astropy2},
SciPy \citep{scipy}.}

\clearpage
\bibliographystyle{aasjournal}
\bibliography{ref.bib}

\begin{appendix}

\section{Virial analysis}
\label{app:virial}

Here we present the equations for calculating each energy term in the Virial equation.
For the \textit{i}th layer in our onion model, we denote 
the H$_2$ number density as $n_{{\rm H_2},i}$, 
the temperature as $T_{{\rm kin},i}$, the inner and outer radii as $r_{i-1}$ and $r_i$, and the midpoint radius of the layer as $r_{i-\frac{1}{2}}=\frac{1}{2}(r_{i-1}+r_i)$, 
where $i=1,...,N$, $r_0=0$, and $r_N=R$. 
The layer width is denoted as $\Delta r$.

We assume the gas is composed of molecular hydrogen, helium, and metals with
the mass fractions of
$X=0.7110$, $Y=0.2741$, and $Z=0.0149$, respectively \citep{Lodders03}.
The total number density, $n$, can be expressed as 
\begin{equation}
n = n_{\rm H_2} + n_{\rm He} + n_{\rm metal}
= \frac{\rho X}{2m_{\rm H}} 
+ \frac{\rho Y}{4m_{\rm H}} 
+ \frac{\rho Z}{15.5m_{\rm H}} 
= \frac{\rho}{m_{\rm H}}
\left(
\frac{X}{2} + \frac{Y}{4} + \frac{Z}{15.5}
\right)
=\frac{\rho}{m_{\rm H}}\frac{1}{\mu},
\end{equation}
where $\rho$ is volume mass density, $m_{\rm H}$ is the mass of the hydrogen atom and 
$\mu$ is the mean molecular weight per particle ($\mu=2.35$).
The mean molecular weight per molecular hydrogen ($\mu_{\rm H_2}=2.81$) can be found by
\begin{equation}
n_{\rm H_2}
= \frac{\rho X}{2m_{\rm H}}
=\frac{\rho}{m_{\rm H}}\frac{1}{\mu_{\rm H_2}}.
\end{equation}
The volume mass density, $\rho$, can be expressed by either the total number density, $n$, or the H$_2$ number density, $n_{\rm H_2}$, with
\begin{align}
\rho
&= \mu n m_{\rm H}\\
&= \mu_{\rm H_2} n_{\rm H_2} m_{\rm H}\label{equ:mass_den}.
\end{align}
The total number density, $n$, can be expressed by the H$_2$ number density, $n_{\rm H_2}$, with
\begin{equation}
n = \frac{\mu_{\rm H_2}}{\mu} n_{\rm H_2}.
\end{equation}
Another widely assumed gas composition is 
$n_{\rm H_2}=5n_{\rm He}$, where metals are negligible \citep[e.g.,][ and also the DCF formula of Equation \ref{equ:DCF}]{Myers83c}.
The corresponding mean molecular weights under this assumption are
$\mu=2.33$ and $\mu_{\rm H_2}=2.8$.
Although we adopt a different but more realistic gas composition in the Virial analysis from the one \citet{Crutcher04} adopted in the DCF formula, the $\mu_{\rm H_2}$ from the two assumptions are similar and thus the derived volume mass density (Equation \ref{equ:mass_den}) would not have any significant difference.

We calculate the gravitational energy by
\begin{align}
E_{\rm grav}
&=-4\pi\int_0^R GM(r)\rho(r)rdr\\
&\approx-4\pi G \mu_{\rm H_2} m_{\rm H} \sum_{i=1}^N
M(r_{i-\frac{1}{2}}) n_{{\rm H_2},i} r_{i-\frac{1}{2}} \Delta r,
\end{align}
where $G$ is the gravitational constant, and $M(r)$ is the fractional core mass inside a radius of $r$.

We calculate the thermal energy by
\begin{align}
E_{\rm thermal}
&=4\pi\int_0^R \frac{3}{2}n(r)k_{\rm B}T_{\rm kin}(r) r^2 dr\\
&\approx 4\pi \frac{3\mu_{\rm H_2}k_{\rm B}}{2\mu}  \sum_{i=1}^N
n_{{\rm H_2},i} T_{{\rm kin}, i} r_{i-\frac{1}{2}}^2 \Delta r,
\end{align}
where $k_{\rm B}$ is the Boltzmann constant.

The surface kinetic term due to external energy, $P_{\rm ext}$, is,
\begin{equation}
E_{\rm surface}
=\frac{3}{2}P_{\rm ext}V=2\pi P_{\rm ext}R^3,
\end{equation}
where $V$ is the core volume.
Here we assume that $P_{\rm ext}=n_{\rm i=N} k_{\rm B} T_{\rm kin,i=N}$ is equal to the thermal pressure in the outermost layer.

In our onion model, the one-dimensional non-thermal velocity dispersion, $\sigma_{v, \rm NT}$, is a constant value of 0.046~km~s$^{-1}$.
The turbulent energy is calculated by
\begin{equation}
E_{\rm turb}
=\frac{3}{2}M\sigma_{v, \rm NT}^2,
\end{equation}
where $M$ is the total core mass.

The rotational energy is calculated by
\begin{equation}
E_{\rm rot}= \frac{1}{2} \sum_{i=1}^{N} I_{\rm i}\omega_{\rm i}^2
=\frac{1}{2} \sum_{i=1}^{N} 
\frac{2}{5}(M(r_i)-M(r_{i-1}))\frac{r_{i}^5-r_{i-1}^5}{r_{i}^3-r_{i-1}^3}\omega_{\rm i}^2,
\end{equation}
where $I_{\rm i}$ and $\omega_{\rm i}$ are the moment of inertia and angular velocity of the $i$th layer, respectively.

The magnetic energy is calculated in the cgs units by
\begin{equation}
E_{\rm mag}= \frac{B^2}{8\pi}V.
\end{equation}

\end{appendix}

\end{document}